\documentclass[fleqn,usenatbib]{mnras}

\usepackage[T1]{fontenc}

\DeclareRobustCommand{\VAN}[3]{#2}
\let\VANthebibliography\thebibliography
\def\thebibliography{\DeclareRobustCommand{\VAN}[3]{##3}\VANthebibliography}

\usepackage{graphicx}	
\usepackage{amsmath}	
\usepackage{amssymb}	
\usepackage{orcidlink}

\usepackage{newtxtext,newtxmath}

\newcommand{\HeI}{He~{\sc i}}
\newcommand{\OI}{O~{\sc i}}
\newcommand{\NaI}{Na~{\sc i}}

\newcommand{\FeII}{Fe~{\sc ii}}
\newcommand{\CaII}{Ca~{\sc ii}}
\newcommand{\SiII}{Si~{\sc ii}}
\newcommand{\ScII}{Sc~{\sc ii}}
\newcommand{\BaII}{Ba~{\sc ii}}

\defcitealias{piro21}{P21}
\defcitealias{arnett82}{A82}
\defcitealias{wheeler15}{W15}

\title[SN~2021bxu]{Fast and Not-so-Furious: Case Study of the Fast and Faint Type IIb \\ SN~2021bxu}

\author[D. D. Desai et al.]{
Dhvanil D. Desai\orcidlink{0000-0002-2164-859X}$^{1}$\thanks{E-mail: dddesai@hawaii.edu},
Chris Ashall\orcidlink{0000-0002-5221-7557}$^{2}$,
Benjamin J. Shappee\orcidlink{0000-0003-4631-1149}$^{1}$,
Nidia Morrell\orcidlink{0000-0003-2535-3091}$^{3}$,
Llu\'is Galbany\orcidlink{0000-0002-1296-6887}$^{4,5}$,
\newauthor 
Christopher R. Burns$^{6}$,
James M. DerKacy\orcidlink{0000-0002-7566-6080}$^{2}$,
Jason T. Hinkle\orcidlink{0000-0001-9668-2920}$^{1}$,
Eric Hsiao$^{7}$,
Sahana Kumar\orcidlink{0000-0001-8367-7591}$^{7}$,
\newauthor 
Jing Lu\orcidlink{0000-0002-3900-1452}$^{7}$,
Mark M. Phillips\orcidlink{0000-0003-2734-0796}$^{3}$,
Melissa Shahbandeh\orcidlink{0000-0002-9301-5302}$^{8}$,
Maximilian D. Stritzinger\orcidlink{0000-0002-5571-1833}$^{9}$,
Eddie Baron$^{10,11}$,
\newauthor 
Melina C. Bersten$^{12,13,14}$,
Peter J. Brown\orcidlink{0000-0001-6272-5507}$^{15}$,
Thomas de Jaeger$^{1}$,
Nancy Elias-Rosa$^{4}$,
Gast\'on Folatelli$^{12,13,14}$,
\newauthor 
Mark E. Huber\orcidlink{0000-0003-1059-9603}$^{1}$,
Paolo Mazzali$^{16}$,
Tom\'as E. M\"uller-Bravo\orcidlink{0000-0003-3939-7167}$^{4,5}$,
Anthony L. Piro$^{6}$,
Abigail Polin$^{6}$,
\newauthor 
Nicholas B. Suntzeff$^{15}$,
Joseph P. Anderson\orcidlink{0000-0003-0227-3451}$^{17,18}$,
Kenneth C. Chambers\orcidlink{0000-0001-6965-7789}$^{1}$,
Ting-Wan Chen\orcidlink{0000-0002-1066-6098}$^{19,20}$,
\newauthor 
Thomas de Boer\orcidlink{0000-0001-5486-2747}$^{1}$,
Michael D. Fulton\orcidlink{0000-0003-1916-0664}$^{21}$,
Hua Gao\orcidlink{0000-0003-1015-5367}$^{1}$,
Mariusz Gromadzki\orcidlink{0000-0002-1650-1518}$^{22}$,
Cosimo Inserra\orcidlink{0000-0002-3968-4409}$^{23}$,
\newauthor 
Eugene A. Magnier\orcidlink{0000-0002-7965-2815}$^{1}$,
Matt Nicholl\orcidlink{0000-0002-2555-3192}$^{21}$,
Fabio Ragosta\orcidlink{0000-0003-2132-3610}$^{24}$,
Richard Wainscoat\orcidlink{0000-0002-1341-0952}$^{1}$,
David R. Young\orcidlink{0000-0002-1229-2499}$^{21}$ \\
$^{1}$ Institute for Astronomy, University of Hawai`i at M\=anoa, 2680 Woodlawn Drive, Honolulu, HI 96822, USA \\
$^{2}$ Department of Physics, Virginia Tech, Blacksburg, VA 24061, USA \\
$^{3}$ Las Campanas Observatory, Carnegie Observatories, Casilla 601, La Serena, Chile \\
$^{4}$ Institute of Space Sciences (ICE, CSIC), Campus UAB, Carrer de Can Magrans, s/n, E-08193 Barcelona, Spain \\
$^{5}$ Institut d'Estudis Espacials de Catalunya (IEEC), Gran Capit\`{a}, 2-4, Edifici Nexus, Desp. 201, E-08034 Barcelona, Spain \\
$^{6}$ The Observatories of the Carnegie Institution for Science, 813 Santa Barbara St., Pasadena, CA 91101, USA \\
$^{7}$ Department of Physics, Florida State University, 77 Chieftan Way, Tallahassee, FL 32306, USA \\
$^{8}$ Space Telescope Science Institute, 3700 San Martin Drive, Baltimore, MD 21218, USA \\
$^{9}$ Department of Physics and Astronomy, Aarhus University, Ny Munkegade, DK-8000 Aarhus C, Denmark \\
$^{10}$ Homer L. Dodge Department of Physics and Astronomy, University of Oklahoma, 440 W. Brooks, Rm 100, Norman, OK 73019, USA \\
$^{11}$ Hamburger Sternwarte, Gojenbergsweg 112, 21029 Hamburg, Germany \\
$^{12}$ Facultad de Ciencias Astron\'omicas y Geof\'isicas, Universidad Nacional de La Plata, Paseo del Bosque S/N, B1900FWA, La Plata, Argentina \\
$^{13}$ Instituto de Astrof\'isica de La Plata (IALP), CCT-CONICET-UNLP, Paseo del Bosque S/N, B1900FWA, La Plata, Argentina \\
$^{14}$ Kavli Institute for the Physics and Mathematics of the Universe (WPI), The University of Tokyo, 5-1-5 Kashiwanoha, Kashiwa, Chiba, 277-8583, Japan \\
$^{15}$ Mitchell Institute for Fundamental Physics and Astronomy, Department of Physics and Astronomy, Texas A\&M University 4242 TAMU, College Station, TX 77845, USA \\
$^{16}$ Astrophysics Research Institute, Liverpool John Moores University, IC2, Liverpool Science Park, 146 Brownlow Hill, Liverpool L3 5RF, UK \\
$^{17}$ European Southern Observatory, Alonso de C\'ordova 3107, Casilla 19, Santiago, Chile \\
$^{18}$ Millennium Institute of Astrophysics MAS, Nuncio Monsenor Sotero Sanz 100, Off. 104, Providencia, Santiago, Chile \\
$^{19}$ Technische Universit{\"a}t M{\"u}nchen, TUM School of Natural Sciences, Physik-Department, James-Franck-Stra{\ss}e 1, 85748 Garching, Germany \\
$^{20}$ Max-Planck-Institut f{\"u}r Astrophysik, Karl-Schwarzschild Stra{\ss}e 1, 85748 Garching, Germany\\
$^{21}$ Astrophysics Research Centre, School of Mathematics and Physics, Queen’s University Belfast, Belfast BT7 1NN, UK \\
$^{22}$ Astronomical Observatory, University of Warsaw, Al. Ujazdowskie 4, 00-478 Warszawa, Poland \\
$^{23}$ Cardiff Hub for Astrophysics Research and Technology, School of Physics \& Astronomy, Cardiff University, Queens Buildings, The Parade, Cardiff, CF24 3AA, UK \\
$^{24}$ Istituto Nazionale di Astrofisica, Viale del Parco Mellini 84, 00136 Rome, Italy
}

\date{Accepted XXX. Received YYY; in original form ZZZ}

\pubyear{2023}

\begin{document}
\label{firstpage}
\pagerange{\pageref{firstpage}--\pageref{lastpage}}
\maketitle

\begin{abstract}
We present photometric and spectroscopic observations and analysis of SN~2021bxu (ATLAS21dov), a low-luminosity, fast-evolving Type IIb supernova (SN). SN~2021bxu is unique, showing a large initial decline in brightness followed by a short plateau phase. With $M_r = -15.93 \pm 0.16\, \mathrm{mag}$ during the plateau, it is at the lower end of the luminosity distribution of stripped-envelope supernovae (SE-SNe) and shows a distinct $\sim$10 day plateau not caused by H- or He-recombination. SN~2021bxu shows line velocities which are at least $\sim1500\,\mathrm{km\,s^{-1}}$ slower than typical SE-SNe. It is photometrically and spectroscopically similar to Type IIb SNe during the photospheric phases of evolution, with similarities to Ca-rich IIb SNe.  We find that the bolometric light curve is best described by a composite model of shock interaction between the ejecta and an envelope of extended material, combined with a typical SN~IIb powered by the radioactive decay of $^{56}$Ni. The best-fit parameters for SN~2021bxu include a $^{56}$Ni mass of $M_{\mathrm{Ni}} = 0.029^{+0.004}_{-0.005}\,\mathrm{M_{\odot}}$, an ejecta mass of $M_{\mathrm{ej}} = 0.61^{+0.06}_{-0.05}\,\mathrm{M_{\odot}}$, and an ejecta kinetic energy of $K_{\mathrm{ej}} = 8.8^{+1.1}_{-1.0} \times 10^{49}\, \mathrm{erg}$. From the fits to the properties of the extended material of Ca-rich IIb SNe we find a trend of decreasing envelope radius with increasing envelope mass. SN~2021bxu has $M_{\mathrm{Ni}}$ on the low end compared to SE-SNe and Ca-rich SNe in the literature, demonstrating that SN~2021bxu-like events are rare explosions in extreme areas of parameter space. The progenitor of SN~2021bxu is likely a low mass He star with an extended envelope.
\end{abstract}

\begin{keywords}
supernovae:general -- supernovae: individual: SN 2021bxu -- stars: massive
\end{keywords}



\section{Introduction}
Core-collapse (CC) supernovae (SNe) mark the explosive ends of the lives of massive stars ($M \gtrsim 8\,\mathrm{M_{\odot}}$) via gravitational collapse of their stellar cores. Some CC SNe occur from progenitors that have lost their outer envelopes and are classified as stripped-envelope supernovae \citep[SE-SNe;][]{clocchiatti96,matheson01}. Optical spectral signatures of H and He are mainly used to distinguish between the different types of SE-SNe \citep[][]{filippenko97}. The lack of H and \SiII\,$\lambda6150$ features defines Type Ib/c SE-SNe. Furthermore, if there is \HeI\,$\lambda 5876$ absorption present, the SN is a Type Ib and if there is weak or no \HeI\,$\lambda 5876$ absorption, the SN is a Type Ic. In addition to these types, SNe that show transient lines of H, making their late-time spectra appear more like SNe~Ib, are classified as Type IIb SNe. 

The exact nature of progenitor scenarios is unclear but different types of SE-SNe may be explained by various mass loss mechanisms in the progenitor star \citep[][]{filippenko94}. SNe~Ic likely result from stars that have lost both their H and He envelopes. SNe~Ib from stars with less extreme stripping, that have lost H envelope. And SNe~IIb result from stars that have lost most of their H envelope, showing weak H lines at early times \citep[e.g.,][]{filippenko97,shivvers17a,prentice17}. It is unclear if there is a continuum between each type of SE-SN or whether this points towards multiple mass-loss mechanisms. Some common proposed mass loss mechanisms which may be responsible for stripping the stellar envelope(s) include (a) outbursts in Luminous Blue Variables \citep[LBVs; e.g.,][]{smith06}, (b) radiation-driven winds \citep[e.g.,][]{heger03,pauldrach12}, (c) envelope stripping due to close binary interactions \citep[e.g.,][]{podsiadlowski93,woosley95,wellstein99,wellstein01,podsiadlowski04,Benvenuto13}, (d) mass-loss in rapidly rotating Be stars \citep[e.g.,][]{massa75,kogure82,owocki06}, or a combination of these. 

The low-mass end of SE-SN progenitors (zero-age main-sequence (ZAMS) mass $\sim8-12\, \mathrm{M_{\odot}}$) is not well understood \citep[e.g.,][]{janka12}. However, it is known that stars in this mass range could end their evolution as white dwarfs, explode as low-luminosity electron-capture SNe, or ignite nuclear burning leading to ``standard'' iron core-collapse explosions \citep[e.g.,][]{Miyaji80,Nomoto84,Nomoto87,Hillebrandt84,Miyaji87,heger03}. Their end stages of evolution depend strongly on composition, metallicity, mass-loss history, etc. \citep[e.g.,][and references therein]{Poelarends08}. In the case of SN explosions from progenitors with $M_{\mathrm{ZAMS}}$ in the range $\sim8-12\, \mathrm{M_{\odot}}$, it is possible to observe (a) relatively faint Type II SNe (IIP or IIL depending on the mass of the H-rich envelope) with presumably low degree of interaction, (b) Type IIb SNe having a stronger interaction with a circumstellar material, or (c) SE-SNe \citep[e.g.,][and references therein]{Pumo09}. Given a stellar initial mass function (IMF) that drops steeply towards higher masses, we should expect a significant fraction of CC SNe to have lower-mass progenitors \citep[e.g.,][]{sukhbold16}. However, due to their expected lower luminosities and rapid photometric evolution, SNe resulting from low-mass stars are likely more difficult to observe and follow up, leading to an observational bias \citep{nomoto82,janka12}.

Recently, Calcium-rich transients have emerged as a new class of SNe showing faster photometric evolution than normal SNe, lower luminosities, and a nebular phase dominated by calcium emission \citep[][]{perets11,kasliwal12,Shen19,das22}. Ca-rich transients consist of two main sub-types: I and II. The Type I Ca-rich transients may come from the explosion of white dwarfs (WDs) or highly-stripped CC events \citep{Kawabata10,Tauris15}. They do not show hydrogen in their spectra and are usually found in the outskirts of early-type galaxies, in old, metal-poor environments \citep[][]{perets11}. The Type II Ca-rich transients, which possibly come from a CC event, show hydrogen features. In particular, Ca-rich Type IIb have spectra similar to SNe IIb near peak light, but rapidly evolve into nebular phase ($\sim$30 days after explosion) and show a [\CaII]/[\OI] ratio of $\geq 2$ \citep[][]{das22}. Contrary to the Ca-rich Type I transients, Ca-rich Type IIb SNe are found in star-forming regions and suggest a new class of strongly-stripped SNe (SS-SNe) which have ejecta masses less than $\sim 1\, \mathrm{M_{\odot}}$ and stripped low-mass He stars as their progenitors.

One of the most promising ways to study the progenitor and its outermost layers is through early-time observations of SN explosions \citep[e.g.,][]{bersten12,Piro13,vallely21}. Shock-breakout is an early-time phenomenon that occurs on a timescale of a few minutes to hours when the shock from a CC explosion of a massive star breaks through the stellar surface \citep{waxman17}. For CC events, shock-breakout is a promising tool for measuring the properties of the exploding star through early-time observations of the outer layers. This manifests as X-ray and ultraviolet (UV) brightening, followed by a post-shock breakout cooling phase where most of the radiation is emitted in UV and optical and the envelope expands while cooling down. The time-scale of this phenomenon depends on the type of progenitor and the presence or absence of extended material around the star. For example, for normal SNe~Ib or Ic this lasts a few hours \citep[e.g.,][]{xiang19} and for SNe~IIb with extended envelopes this is on the order of a few days \citep[e.g.,][]{nomoto93}. Since the shock-breakout and cooling depend on properties of the outermost layers of the progenitor, observing those via early-time observations can provide measurements of the temperature, radius, and mass of the envelope or circumstellar material \citep[e.g.,][]{Nakar14,piro21}.

Few SE-SNe have been observed to exhibit so-called "double-peaked" light curves with an initial decline due to shock cooling and a second peak due to the radioactive decay of $^{56}$Ni, which normally powers the light curve. Generally, these objects were classified as SNe~IIb. The first of this class was the well-studied SN~1993J \citep[][]{richmond94}, followed by other SE-SNe, e.g., SN~2011dh \citep{arcavi11}, SN~2011fu \citep{moralesgaroffolo15}, SN~2013df \citep{moralesgaroffolo14}, and some others as shown in \citet{prentice20}. The initial decline in SN~1993J was explained as a result of a lengthened shock cooling due to the shock passing through an extended envelope around the star instead of ``breaking out'' at an abrupt surface, producing a longer initial decline. Similar to that observed in various SNe~IIb, a subset of Ca-rich transients have also been shown to have double-peaked light curves with an initial decline of a few days that are well fit by an extended envelope model \citep[][]{das22}. For example, \citet{Ertini23} showed that SN~2021gno, a Ca-rich SN Ib with double-peaked light curves, can be well modelled by a CC explosion of a highly-stripped, massive star.

Exotic classes of SNe are being discovered by untargeted surveys such as the All-Sky Automated Survey for SuperNovae \citep[ASAS-SN;][]{shappee14,kochanek17}, the Asteroid Terrestrial-impact Last Alert System \citep[ATLAS;][]{tonry18,smith20} and the Zwicky Transient Facility \citep[ZTF;][]{masci19,bellm19}, and studied with high-precision multi-band early follow-up by programs such as the Precision Observations of Infant Supernova Explosions \citep[POISE;][]{burns21} collaboration. SN~2021csp \citep[][]{fraser21}, SN~2021fxy \citep[][]{DerKacy22}, SN~2021gno \citep[][]{Ertini23} and SN~2021aefx \citep[][]{ashall22} are all SNe followed-up by POISE, and each of them demonstrates the power of obtaining high-precision early-time observations by providing insight into their sub-class of SNe. 

In this work we study SN 2021bxu\footnote{\url{https://www.wis-tns.org/object/2021bxu}} (ATLAS21dov), a peculiar SE-SN discovered by ATLAS on UT 06.3 Feb 2021 \citep[][]{tonry21} and later classified as a Type IIb SN \citep[][]{derkacy21}. Early-time observations from POISE show a fast declining light curve followed by a $\sim$10 day plateau and an unusually low peak luminosity. Here we present a detailed study of this unique event with high-precision photometry and spectroscopy, explore a new parameter space for SNe and their progenitors, and derive physical quantities and compare them with other known SE-SNe.

In Section~\ref{sec:host}, we provide the properties of the host galaxy. In Section~\ref{sec:data}, we describe the photometric and spectroscopic observations of SN~2021bxu. In Section~\ref{sec:phot_anal}, we analyze and compare the multi-band light curves and colour curves of SN~2021bxu with SE-SNe and Ca-rich IIb SNe from literature. We also derive and compare the pseudo-bolometric and bolometric light curves of SN~2021bxu with a sample of SE-SNe. In Section~\ref{sec:spec_anal}, we identify the observed spectroscopic lines, compare the line velocities to a sample of SE-SNe and Ca-rich IIb SNe, estimate the time of explosion, and compare SN~2021bxu with similar SNe. In Section~\ref{sec:modelling}, we outline the models used to fit the bolometric and pseudo-bolometric light curves, describe the fitting method and present the best-fit results for the explosion parameters. In Section~\ref{sec:discussion}, we compare and contextualize the physical parameters of SN~2021bxu with a sample of SE-SNe and Ca-rich IIb SNe. Finally, in Section~\ref{sec:conclusion}, we summarize our work and present the conclusions.

\section{Host Galaxy Properties}
\label{sec:host}
The host galaxy ESO~478-~G~006 is classified as an Sbc galaxy with luminosity class II-III \citep{deVaucouleurs91}. SN~2021bxu was discovered in one of the spiral arms of its host galaxy. 

\begin{figure}
	\includegraphics[width=\columnwidth]{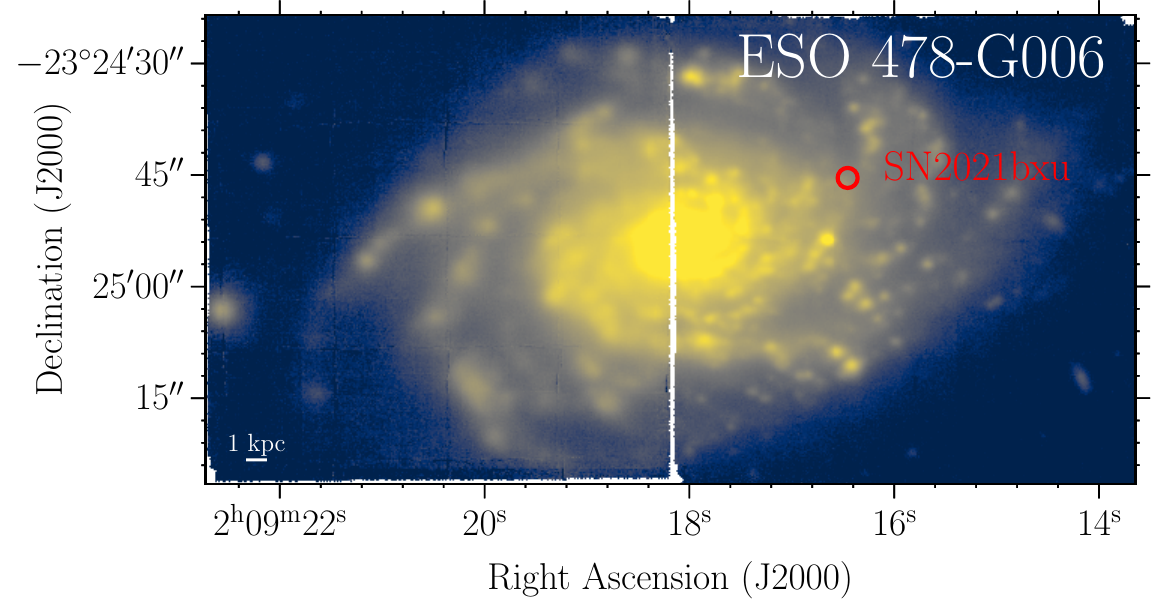}
    \caption{Finding chart of SN~2021bxu in its host galaxy ESO~478-~G~006 obtained with MUSE. The background image is a stellar-continuum subtracted H$\alpha$ emission map obtained from the two MUSE pointings. The different conditions of the observations are evident, the west half of the galaxy where SN~2021bxu exploded presents a better spatial resolution ($0\farcs72$ seeing) than the east half ($2\farcs01$ seeing).}
    \label{fig:host}
\end{figure}

Using the Fitting and Assessment of Synthetic Templates \citep[FAST;][]{kriek09}, we fit stellar population synthesis models to the archival photometry (see Table~\ref{tab:host_phot} in Appendix~\ref{app:phot_tables}) of the host galaxy ESO~478-~G~006 to obtain the global age, the total stellar mass, the star formation rate (SFR), and the specific SFR (sSFR). These global parameters are reported in Table~\ref{tab:host_properties} as best-fit values with uncertainties from Monte Carlo. Our fit assumes a \citet{cardelli89} extinction law with $R_{V} = 3.1$, a Salpeter IMF \citep{salpeter55}, an exponentially declining star-formation rate, the \citet{bruzual03} stellar population models, and allows for a variable average host galaxy extinction higher than the Galactic foreground value of $A_{V,\mathrm{MW}} = 0.045\, \mathrm{mag}$. The value of host extinction from this fit indicates an average across the entire galaxy and is not necessarily the value at the site of the SN. 

To estimate the local host extinction and other properties, we used integral field spectroscopy (IFS) obtained with MUSE mounted to the 8.2m Very Large Telescope (VLT) in October 2016 as a part of the All-weather MUse Supernova Integral-field of Nearby Galaxies \citep[AMUSING;][]{Galbany16,lopez-coba20}) survey. 
Two pointings were observed covering the position of SN~2021bxu (see Figure~\ref{fig:host}). 
Following previous analysis of IFS data \citep[e.g.,][]{Galbany14,Galbany18}), we extracted a $2\arcsec$ aperture spectrum from the IFS cube centered at the SN location, and performed spectral synthesis with STARLIGHT \citep{CidFernandes05}.
By subtracting the best STARLIGHT fit from the observed spectrum, we get the gas-phase spectrum, where we fit Gaussian profiles to the strongest emission lines to measure the main properties of the ionized gas.
In particular, the dust extinction along the line of sight is estimated using the colour excess ($E(B-V)$) from the ratio of the Balmer lines, assuming an intrinsic ratio $I(\mathrm{H}\alpha)/I(\mathrm{H}\beta)=2.86$, valid for case B recombination with $T=10{,}000\,\mathrm{K}$ and electron density $10^2\,\mathrm{cm^{-3}}$ \citep{Osterbrock06}, and using a \cite{cardelli89} extinction law.
Local properties are also listed in Table~\ref{tab:host_properties}.

\begin{table}
	\centering
	\caption{Global and Local Properties of the Host Galaxy ESO~478-~G~006}
	\label{tab:host_properties}
	\renewcommand{\arraystretch}{1.5}
	\begin{tabular}{lcc}
		\hline
            \hline
                & Global (phot) & Local (spec) \\
		\hline
		Age [yr]                             & $2.2^{+8.5}_{-1.2} \times 10^{8}$  & $9^{+11}_{-8} \times 10^{8}$ \\
		$M_*$ [$M_{\odot}$]                  & $2.3^{+1.3}_{-0.6} \times 10^{10}$ & $(3.2\pm0.2) \times 10^{6}$ \\
		SFR [$\mathrm{M_{\odot}\, yr^{-1}}$] & $34^{+16}_{-27}$                   & $(7.38 \pm 0.09)\times 10^{-5}$ \\ 
		sSFR [$\mathrm{yr^{-1}}$]            & $1.5^{+1.2}_{-1.3} \times 10^{-9}$ & $2.3^{+1.2}_{-0.4} \times 10^{-11}$ \\
		$A_{V\mathrm{,host}}$ [mag]          & $1.4^{+0.2}_{-0.9}$                & $0.4\pm0.2$ \\
            $12 + \log_{10}$(O/H)                & $-$                                & $8.71\pm0.14$\\
		\hline
	\end{tabular}
\end{table}

For the purposes of magnitude corrections for SN~2021bxu, we use a null line-of-sight host extinction at the SN site, as inferred from the absence of strong narrow \NaI\,D line in the SN spectra. We do not use $A_{V,\mathrm{host}} = 0.4 \pm 0.2\,\mathrm{mag}$ from local spectroscopic analysis of the host galaxy because this measurement is of the extinction from the total line-of-sight dust column. Since we do not know if the SN exploded in front of the dust column or behind it, using the galaxy measurement is not optimal. However, we note that there may be up to $0.4\,\mathrm{mag}$ of host extinction and including this in the analysis does not change the conclusions of this study.

\section{Data} \label{sec:data}
\subsection{Photometry}
\begin{table}
	\centering
	\caption{Properties of SN~2021bxu}
	\label{tab:21bxu_properties}
	\begin{tabular}{lr} 
		\hline
		RA (J2000) & 02:09:16.47\\
        $\delta$ (J2000) & $-23$:24:45.15\\
        $z$ $^{\mathrm{a}}$& $0.0178$\\
        Host Galaxy & ESO~478-~G~006\\
        Host Offset $^{\mathrm{b}}$ & $9.2 \pm 0.6\,\mathrm{kpc}$\\
        $E(B - V)_{\mathrm{MW}}$ $^{\mathrm{c}}$ & $0.014 \, \mathrm{mag}$\\
        $E(B - V)_{\mathrm{Host}}$ & $\sim0 \, \mathrm{mag}$\\
        $\mu$ $^{\mathrm{a}}$ & $34.28 \pm 0.16 \, \mathrm{mag}$\\
        $D_\mathrm{L}$ $^{\mathrm{d}}$ & $72 \pm 5 \, \mathrm{Mpc}$\\
        Last Non-Detection (MJD) & $59245.12\, \mathrm{days}$\\
        Discovery (MJD) & $59251.28\, \mathrm{days}$\\
        Estimated Explosion (MJD) $^{\mathrm{e}}$ & $59246.3\, \pm 0.4\, \mathrm{days}$\\
		\hline
	\end{tabular}\\
	\begin{flushleft}
	$^\mathrm{a}$\citet{scolnic18}. \\
	$^\mathrm{b}$Projected offset from the host-galaxy nucleus. \\ 
        $^\mathrm{c}$\citet{schlafly11}. \\ 
	$^\mathrm{d}$Computed using $\mu$ from \citet{scolnic18}. \\
        $^\mathrm{e}$See Section~\ref{subsec:t_exp}.
	\end{flushleft}
\end{table}

We present high-precision multi-band photometry of SN~2021bxu obtained with the Henrietta Swope 1.0 m telescope at Las Campanas Observatory, ASAS-SN, ATLAS, and the Panoramic Survey Telescope \& Rapid Response System \citep[Pan-STARRS;][see Table~\ref{tab:bxu_phot} in Appendix~\ref{app:phot_tables}]{flewelling20,chambers16}. 

The Swope photometry, in $BVugri$ bands, was produced using custom reduction and calibration procedures as described in \citet{krisciunas17} and \citet{phillips19} via the POISE collaboration, which builds on the legacy of the Carnegie Supernova Project \citep[CSP;][]{phillips19}. The science images were host-galaxy template subtracted and the nightly zero-points were obtained on photometric nights by observing photometric standards from the \citet{Landolt92} and \citet{Smith02} catalogs. Using these zero-points, we computed natural magnitudes of the local sequence stars (listed in Table~\ref{tab:local_seq} in Appendix~\ref{app:phot_tables}) in the field, which is then used to calibrate the Swope photometry of SN~2021bxu. The Swope photometry is ultimately in the CSP natural system. The ATLAS photometry, in $o$ and $c$ bands, was obtained from the ATLAS Forced Photometry server\footnote{\url{https://fallingstar-data.com/forcedphot/}} \citep{Shingles21}, with photometry produced as outlined in \citet{tonry18} and \citet{smith20}. The ASAS-SN $g$-band light curve was produced using subtracted aperture photometry from ASAS-SN Sky Patrol\footnote{\url{https://asas-sn.osu.edu/}}. Finally, the Pan-STARRS (PS) observations were taken with both 1.8\,m telescope units located at the summit of Haleakala \citep{chambers16}, in an SDSS-like filter system, denoted as $grizy_{\mathrm{PS}}$, and a broad $w_{\mathrm{PS}}$ filter, which is a composite of the $gri_{\mathrm{PS}}$ filters. Pan-STARRS data are processed in real-time as described in \cite{Magnier20a,Magnier20b} and \cite{Waters20}. The data are subject to difference imaging with the Pan-STARRS1 $3\pi$ sky survey data \citep{chambers16} used as references, and photometric zero-points on the target images were set with field stars from the Pan-STARRS1 $3\pi$ catalogue \citep{flewelling20}.

\begin{figure}
	\includegraphics[width=\columnwidth]{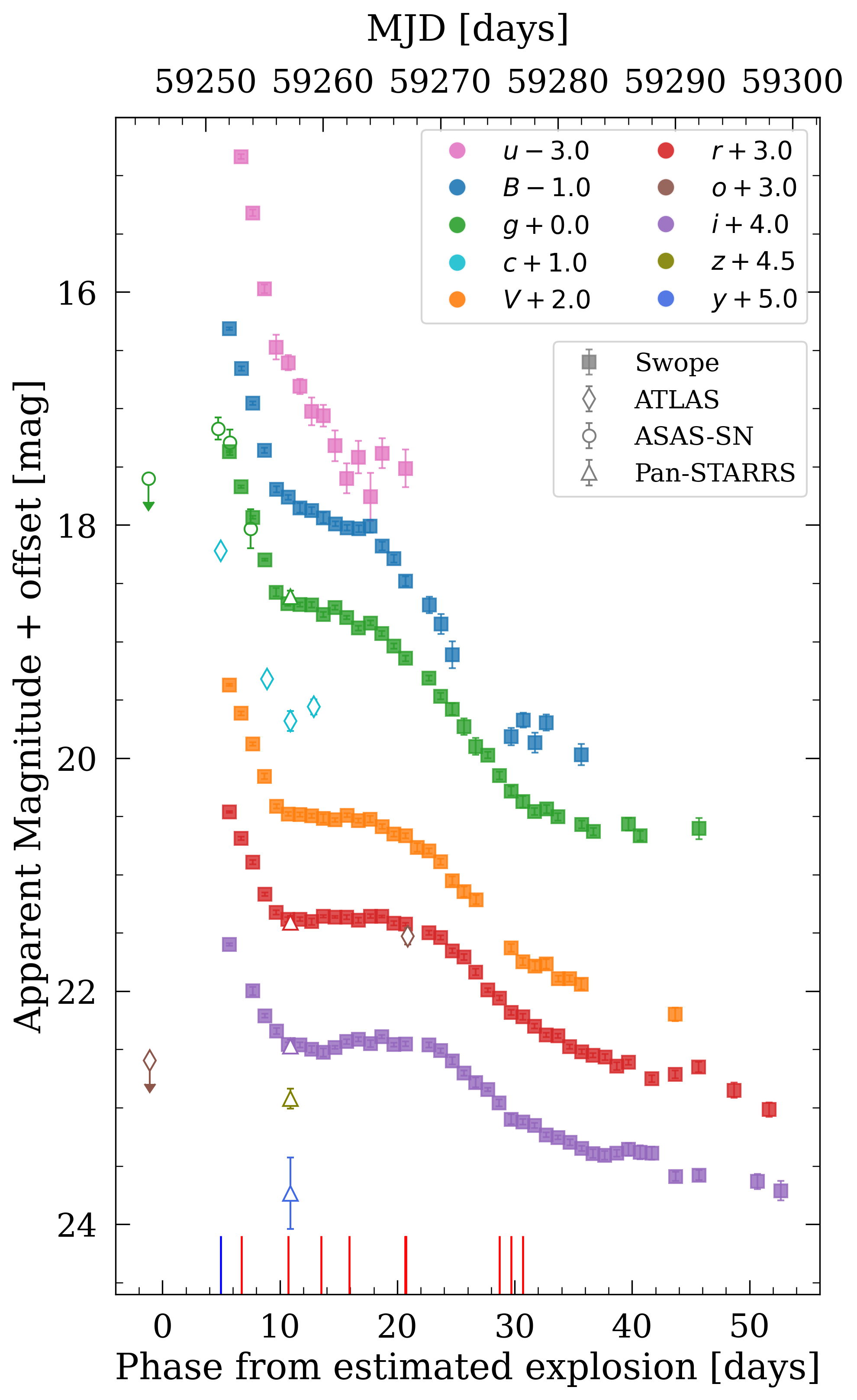}
    \caption{Multi-band light curves of SN~2021bxu from Swope $BVugri$, ATLAS $oc$, ASAS-SN $g$ and Pan-STARRS $grizy$. Non-detections in ASAS-SN $g$ and ATLAS $o$ are shown as points with downward-facing arrows. Vertical red line-segments at the bottom of the plot mark the epochs of spectroscopic observations and the blue line-segment marks the discovery. The estimated time of explosion is further explained in Section~\ref{subsec:bol-lc}.}
    \label{fig:lc}
\end{figure}

All light curves from Swope, ATLAS, ASAS-SN, and Pan-STARRS are shown in Figure~\ref{fig:lc}. The Pan-STARRS $w$ band has a non-detection with a $5\sigma$ upper limit of $m_{w_{\mathrm{PS}}} > 22.3\, \mathrm{mag}$ 33 days before discovery showing no previous outbursts and an upper limit of $m_{w_{\mathrm{PS}}} > 21.7\, \mathrm{mag}$ 216 days after the last measurement from Swope. The ATLAS $o$ band has a non-detection with a $5\sigma$ upper limit of $m_o > 19.6\, \mathrm{mag}$ 6.02 days before discovery and an upper limit of $m_o > 19.8\, \mathrm{mag}$ 101 days after the last measurement from Swope. ASAS-SN also has a non-detection with a $5\sigma$ upper limit of $m_g > 17.6\, \mathrm{mag}$ 6.16 days before discovery along with the first detection at a maximum of $m_g = 17.17 \pm 0.09\, \mathrm{mag}$ 0.2 days before discovery. The non-detection and the first detection can help constrain the time of explosion (see Section~\ref{subsec:t_exp}). 

We obtained UV observations from the Ultra-Violet Optical Telescope \citep[UVOT;][]{Roming05} on the Neil Gehrels Swift Observatory \citep[][]{gehrels04} about 20 days after the estimated explosion but did not detect the SN. Pre-explosion imaging from 2018 and 2019 is available in the $U$ and $UVW1$ filters because of the Swift Gravitational Wave Galaxy Survey, the intent of which is to obtain galaxy template images before the detection of transients \citep{Klingler19}. Using the pipeline from the Swift Optical Ultraviolet Supernova Archive \citep[SOUSA;][]{Brown14}, we measure upper limits on MJD 59266.3 of $UVW1>19.1\,\mathrm{mag}$ and $U>19.1\,\mathrm{mag}$. These magnitude limits are in the UVOT/Vega system using zero points from \citet{Breeveld11}, the time-dependent sensitivity correction from September 2020\footnote{\url{https://heasarc.gsfc.nasa.gov/docs/heasarc/caldb/swift/docs/uvot/uvotcaldb\_throughput\_06.pdf}}, and an aperture correction updated in 2022.  Subsequent observations over the next 10 days yield similar limits and are available from the SOUSA.

There are no data from ZTF or the Transiting Exoplanet Survey Satellite \citep[TESS;][]{ricker16,fausnaugh21,Fausnaugh22} for this object during the time period of interest.

To convert from apparent to absolute magnitudes we need the distance modulus and extinction corrections. The distance modulus for SN~2021bxu ($\mu = 34.28 \pm 0.16 \, \mathrm{mag}$) is measured using precise redshift-independent distance measurement using the Type Ia SN 2009le that exploded in the host galaxy (ESO~478-~G~006; \citealt{scolnic18}). We infer no host-galaxy extinction at the site of the SN due to a lack of narrow \NaI\, D lines in the SN spectra \citep[][]{phillips13}. Therefore, using $\mu$ and the Galactic extinction correction \citep[$E(B - V)_{\mathrm{MW}} = 0.014 \, \mathrm{mag}$;][]{schlafly11}, we obtain the absolute magnitudes listed in Table~\ref{tab:bxu_phot}. Some of the properties of SN~2021bxu are listed in Table~\ref{tab:21bxu_properties} and the photometric analysis is further discussed in Section \ref{sec:phot_anal}.

\subsection{Spectroscopy}
\begin{figure*}
	\includegraphics[width=\textwidth]{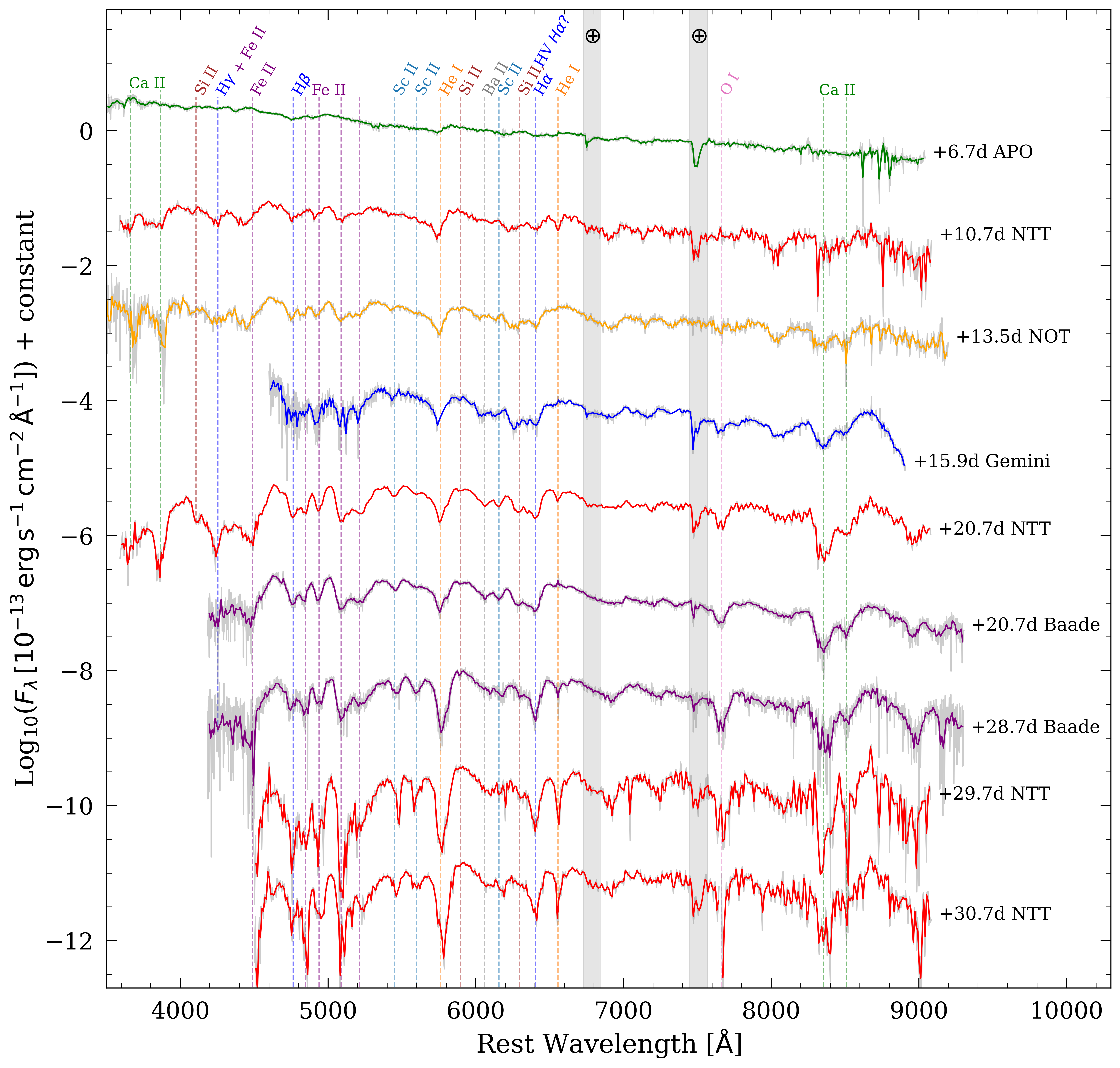}
    \caption{Spectra of SN~2021bxu from APO (green), NTT (red), NOT (orange), Gemini (blue), and Baade (purple) from +6.74 to +30.68 days after its estimated explosion (MJD 59246.3), all listed in Table~\ref{tab:21bxu_spectra}. The spectrum from NTT at +20.7 days is an average of two spectra taken on the same night using the same instrument only half-hour apart.}
    \label{fig:21bxu_spectra}
\end{figure*}

\begin{table}
	\centering
    \caption{Spectroscopy of SN~2021bxu}
	\label{tab:21bxu_spectra}
	\begin{tabular}{cccccc}
		\hline
		\hline
		Date & MJD & Phase$^\mathrm{a}$ & Telescope & Instrument \\
		(UT) & (days) & (days) & &  \\
		\hline
		08 Feb 2021 & 59253.07 & 6.7  & APO    & DIS      \\
		12 Feb 2021 & 59257.02 & 10.7 & NTT    & EFOSC2   \\
		14 Feb 2021 & 59259.84 & 13.5 & NOT    & ALFOSC   \\ 
		17 Feb 2021 & 59262.22 & 15.9 & Gemini & GMOS-N   \\ 
		22 Feb 2021 & 59267.01 & 20.7 & NTT    & EFOSC2   \\ 
		22 Feb 2021 & 59267.04 & 20.7 & NTT    & EFOSC2   \\
		22 Feb 2021 & 59267.05 & 20.7 & Baade  & IMACS    \\ 
		02 Mar 2021 & 59275.04 & 28.7 & Baade  & IMACS    \\
		03 Mar 2021 & 59276.01 & 29.7 & NTT    & EFOSC2   \\
		04 Mar 2021 & 59277.00 & 30.7 & NTT    & EFOSC2   \\
		\hline
	\end{tabular}\\
	\begin{flushleft}
	$^\mathrm{a}$From estimated explosion date (MJD 59246.3) in rest frame.
	\end{flushleft}
\end{table}

Along with the photometry, we also have a total of ten spectroscopic observations for SN~2021bxu. These include optical spectra with spectral range of roughly $4000 - 9000$\,\AA\, obtained from the Dual Imaging Spectrograph (DIS) on the Apache Point Observatory (APO), the ESO Faint Object Spectrograph and Camera v.2 \citep[EFOSC2;][]{Buzzoni84} on the New Technology Telescope (NTT), the Alhambra Faint Object Spectrograph and Camera (ALFOSC) on the Nordic Optical Telescope (NOT), the Gemini Multi-Object Spectrograph \citep[GMOS;][]{Hook04} on the Gemini North telescope, and the Inamori-Magellan Areal Camera and Spectrograph \citep[IMACS;][]{Dressler11} on the Magellan Baade telescope. The spectra from APO and Baade were reduced using standard \textsc{iraf}\footnote{The Image Reduction and Analysis Facility (\textsc{iraf}) is distributed by the National Optical Astronomy Observatory, which is operated by the Association of Universities for Research in Astronomy, Inc., under cooperative agreement with the National Science Foundation.} packages with the methods as described in \citet{hamuy06} and \citet{folatelli13}. The NOT data were taken as a part of the NUTS2 collaboration\footnote{\url{https://nuts.sn.ie/}} and reduced using the \textsc{foscgui} pipeline\footnote{\textsc{foscgui} is a Python-based graphic user interface (GUI) developed by E. Cappellaro and aimed at extracting SN spectroscopy and photometry obtained with FOSC-like instruments. A package description can be found at \url{https://sngroup.oapd.inaf.it/foscgui.html}}. The NTT spectra were obtained through the Public European Southern Observatory Spectroscopic Survey of Transient Objects (PESSTO) program, and reduced using the data reduction pipeline  described in \citet{smartt15}. The Gemini spectra were reduced using a custom-made \textsc{iraf} routine. Dates and phases of all spectra are listed in Table~\ref{tab:21bxu_spectra}.

Figure~\ref{fig:21bxu_spectra} shows the spectral sequence. The original unbinned spectra are in gray and the higher signal-to-noise ratio (SNR), binned spectra are in colours. The unbinned spectra are resampled at a resolution of $10$\,\AA\, using \texttt{SpectRes} \citep{Carnall17} to produce the binned spectra. The two NTT spectra on UT 22 Feb 2021 are taken only one half-hour apart and, since they are on the same telescope and instrument, we average them to produce a combined spectrum with a higher SNR. This combined spectrum is used for all following analysis.

\section{Photometric Analyses} \label{sec:phot_anal}
\subsection{Multi-band Light Curves} \label{subsec:multiband-lc}
\begin{figure*}
	\includegraphics[width=\textwidth]{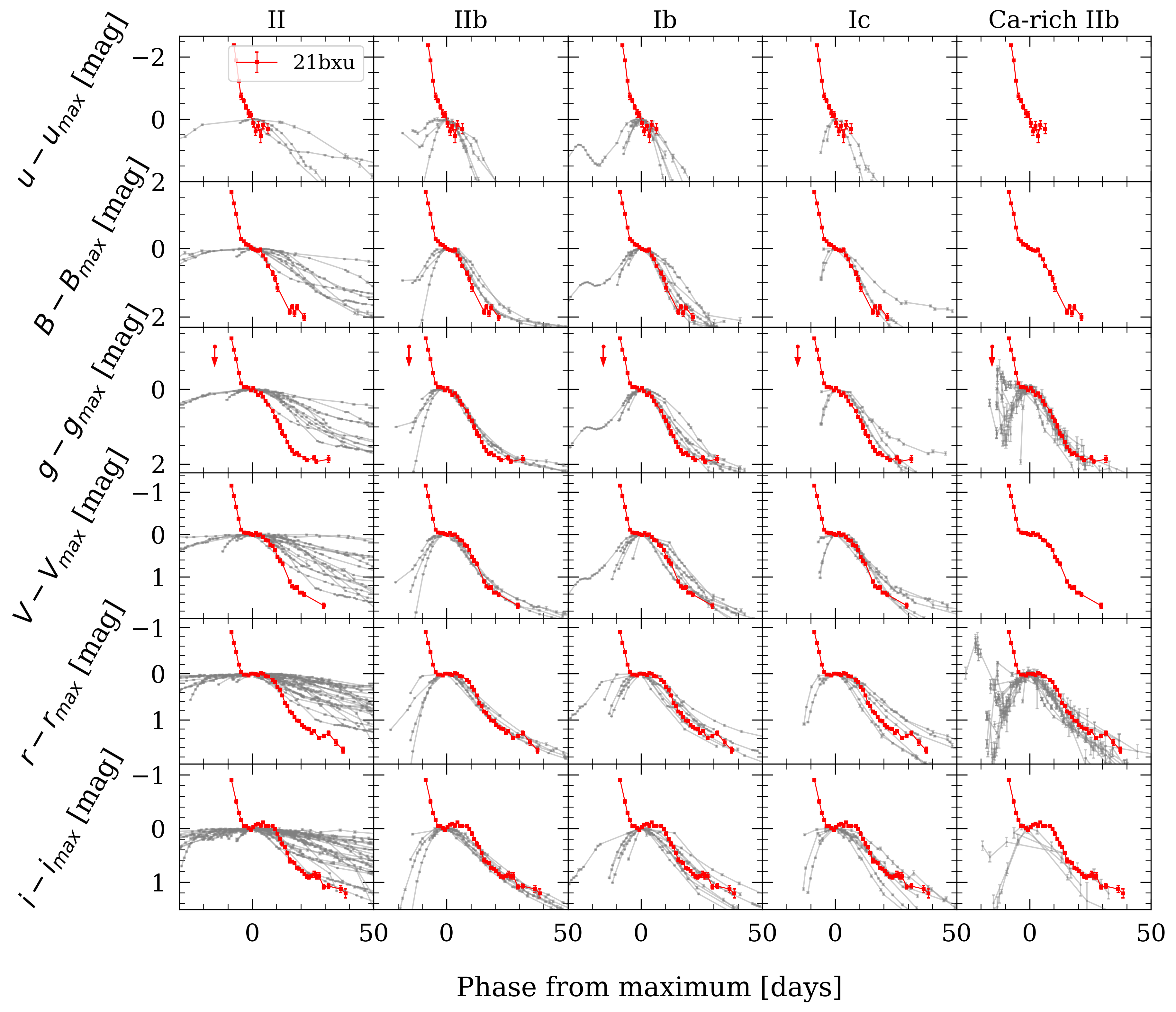}
    \caption{The gray light curves show SNe II, IIb, Ib, and Ic from the Carnegie Supernova Project \citep[][]{stritzinger18a,anderson14} and Ca-rich Type IIb SNe from \citet[][]{das22}. The red light curves in each subplot are SN~2021bxu. The $g$-band panel shows the last non-detection from ASAS-SN as the downward-pointing arrow. The epoch of maximum light for normalization is chosen as the peak of the $^{56}$Ni component of the light curve (see Section~\ref{sec:modelling}).}
    \label{fig:lc_comparison}
\end{figure*}

The multi-band light curves of SN~2021bxu are presented in Figure~\ref{fig:lc}, ranging from the $u$ band to the $i$ band. There is a decrease of $\sim$1.7\,mag in the $u$ band from $\sim$17.9 to $\sim$19.5\,mag in the first 3\,days and thereafter it keeps decreasing almost linearly, but with a shallower slope. However, moving to the optical bands, a plateau starts appearing, which becomes more prominent in the redder bands. For example, in the $g$ band, the brightness declines by $\sim$1.5\,mag in the first $\sim$5\,days after discovery and it is followed by a plateau where the brightness stays roughly constant for the next $\sim$10\,days before declining again. Looking at the $i$ band, where it gets moderately bright again, this plateau may be interpreted as a second peak. Using the distance modulus $\mu = 34.28 \pm 0.16 \, \mathrm{mag}$ from Table~\ref{tab:21bxu_properties}, we obtain a peak $r$-band magnitude of $M_r = -16.86 \pm 0.16 \, \mathrm{mag}$ at the first epoch. After the initial decline, the absolute magnitude during the plateau phase is $M_r = -15.93 \pm 0.16\, \mathrm{mag}$ as estimated from the $^{56}$Ni peak (see Section~\ref{sec:modelling}).

In Figure~\ref{fig:lc_comparison}, we show the light curves of SN~2021bxu in $BVugri$ bands normalized to the bolometric $^{56}$Ni peak (discussed in Section~\ref{sec:modelling}) in each band along with a sample of various types of SE-SNe (IIb, Ib, and Ic) from \citet[][]{stritzinger18a}, a sample of SNe~II from \citet[][]{anderson14}, and a sample of Ca-rich Type IIb SNe from \citet[][]{das22}. The most obvious distinction between SN~2021bxu and the rest of the sample is the presence of a strong initial decline in brightness. Although dissimilar to most SNe~II having the typical long plateau phase, SN~2021bxu shows similarities to the light curves of Type II-L SN~2001fa and SN~2007fz \citep{Faran14} including an initial decline and rise to a second peak. However, SN~2021bxu has much less H in the spectra than the two SNe~II-L. SN~2001fa and SN~2007fz have been proposed as a link between SNe~II and SNe~IIb because of the intermediate strength H$\alpha$ P-Cygni profile between those SN Types. Nevertheless, \citet{Pessi19} finds no evidence of continuum between SNe~II and SNe~IIb. On the other hand, the overall shape of SN~2021bxu's later decline matches well with the SE-SNe and especially well with the Ca-rich IIb SNe and with SNe~IIb. The Ca-rich IIb SNe show a wide range of initial declines depending on the properties of the external layers of the progenitor \citep[][]{das22}. However, the later decline in the $g$ band for SN~2021bxu is almost identical to that of the Ca-rich IIb and SNe~IIb samples.

\subsection{Colour Curves} \label{subsec:cc}
\begin{figure}
	\includegraphics[width=\columnwidth]{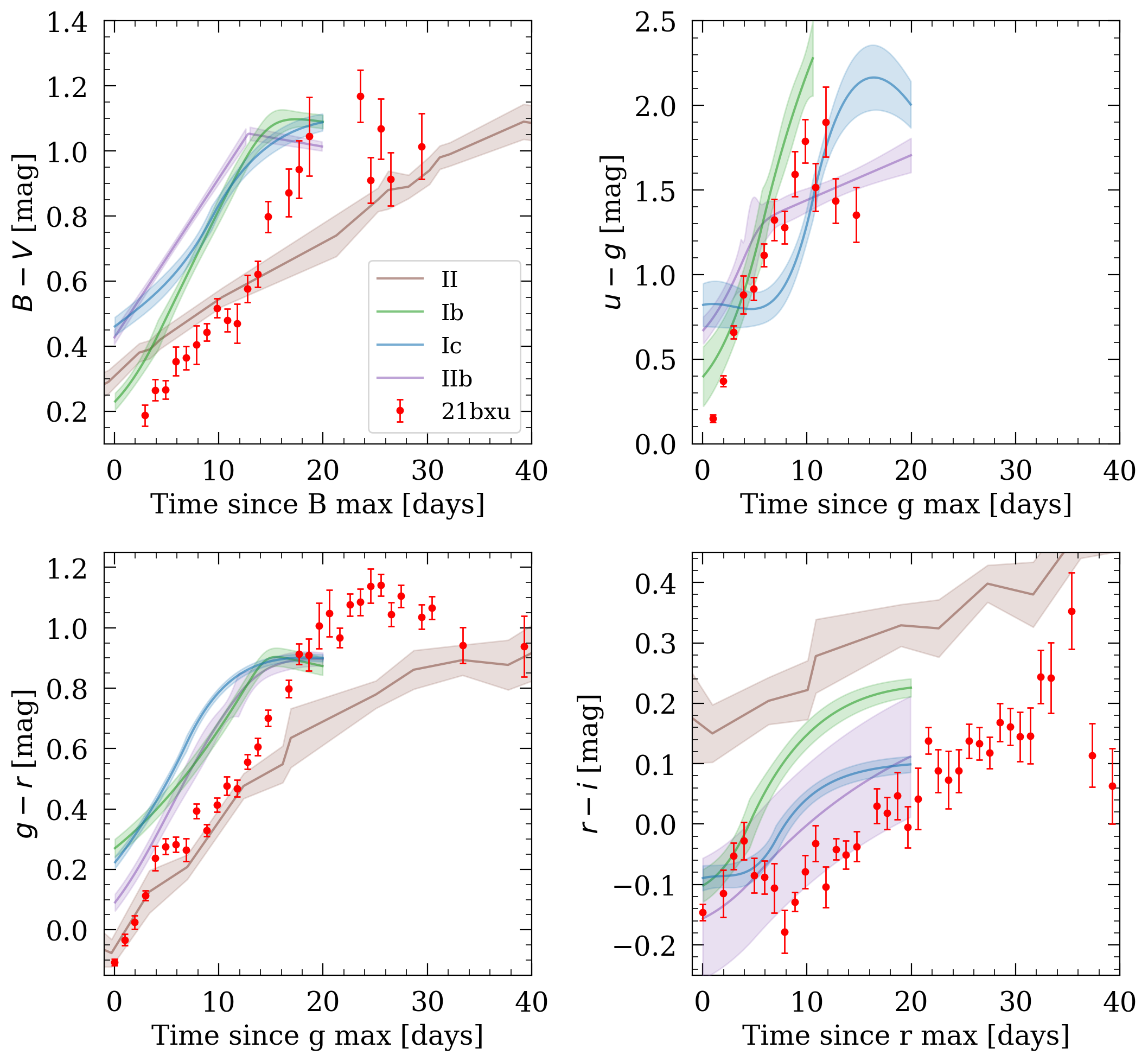}
    \caption{Colour curves of SN~2021bxu using Swope photometry compared to the Type Ib, Ic, and IIb colour curve templates from CSP SE-SNe sample. We create a template for $B-V$ colour of a Type II SN using SN 2014G \citep[][]{bose16} and for $g-r$ and $r-i$ colours of a Type II SN using DES15E1iuh \citep[][]{dejaeger20}. There were no $u$-band data available for Type II SNe, so we do not have a Type II template for the $u-g$ colour. We do not plot colour curves for Ca-rich IIb SNe due to the lack of a statistical sample or a representative SN.}
    \label{fig:cc_comparison}
\end{figure}

With the available multi-band photometry, we produce the Milky Way reddening corrected $B-V$, $u-g$, $g-r$, and $r-i$ colour curves for SN~2021bxu, presented in Figure~\ref{fig:cc_comparison}. The $u-g$ colour quickly reaches a reddest value of $\sim$1.7\,mag $\sim$15\,days after explosion. The $B-V$ and $g-r$ colours follow a trend similar to each other and get to their reddest value of $\sim$1.1\,mag $\sim$25\,days after explosion. The $B-V$ colour shows an unexpected change of slope around $\sim$16-19\,days after explosion corresponding to the time of the plateau phase. On the other hand, starting off at the bluest colour, the $r-i$ colour shows a slow but steady increase from $\sim-$0.1\,mag to $\sim$0.3\,mag over the $\sim$40\,days after explosion. 

We compare the colour curves of SN~2021bxu to templates from \citet[][]{stritzinger18b} for SE-SNe~Ib, Ic, and IIb as well as SNe~II. Figure~\ref{fig:cc_comparison} shows the colour curves of SN~2021bxu together with the templates. The $B-V$ colour of SN~2021bxu is consistent with SNe~Ib until $\sim$5\,days after $B$-band maximum but it is bluer by $\sim$0.4\,mag around day 12. The $u-g$ colour follows SNe~Ib up to $\sim$5\,days after $g$-band maximum, but resembles SNe~IIb thereafter. On the other hand, the $g-r$ colour does not seem to match any of the types except the first $\sim$3\,days where it matches with SNe~II; it is too blue for SE-SNe and too red for SNe~II after that. The $B-V$ and the $g-r$ colours peak $\sim$10 days after the templates peak, although the templates only go up to 20 days past maximum. Finally, the $r-i$ colour matches the trend of SNe~IIb given the error bars on the data and the model. Although SN~2021bxu shows a plateau in its light curve, the colour curves do not match those of SNe~II. 

\subsection{Bolometric Light Curves} \label{subsec:bol-lc}
\begin{figure}
	\includegraphics[width=\columnwidth]{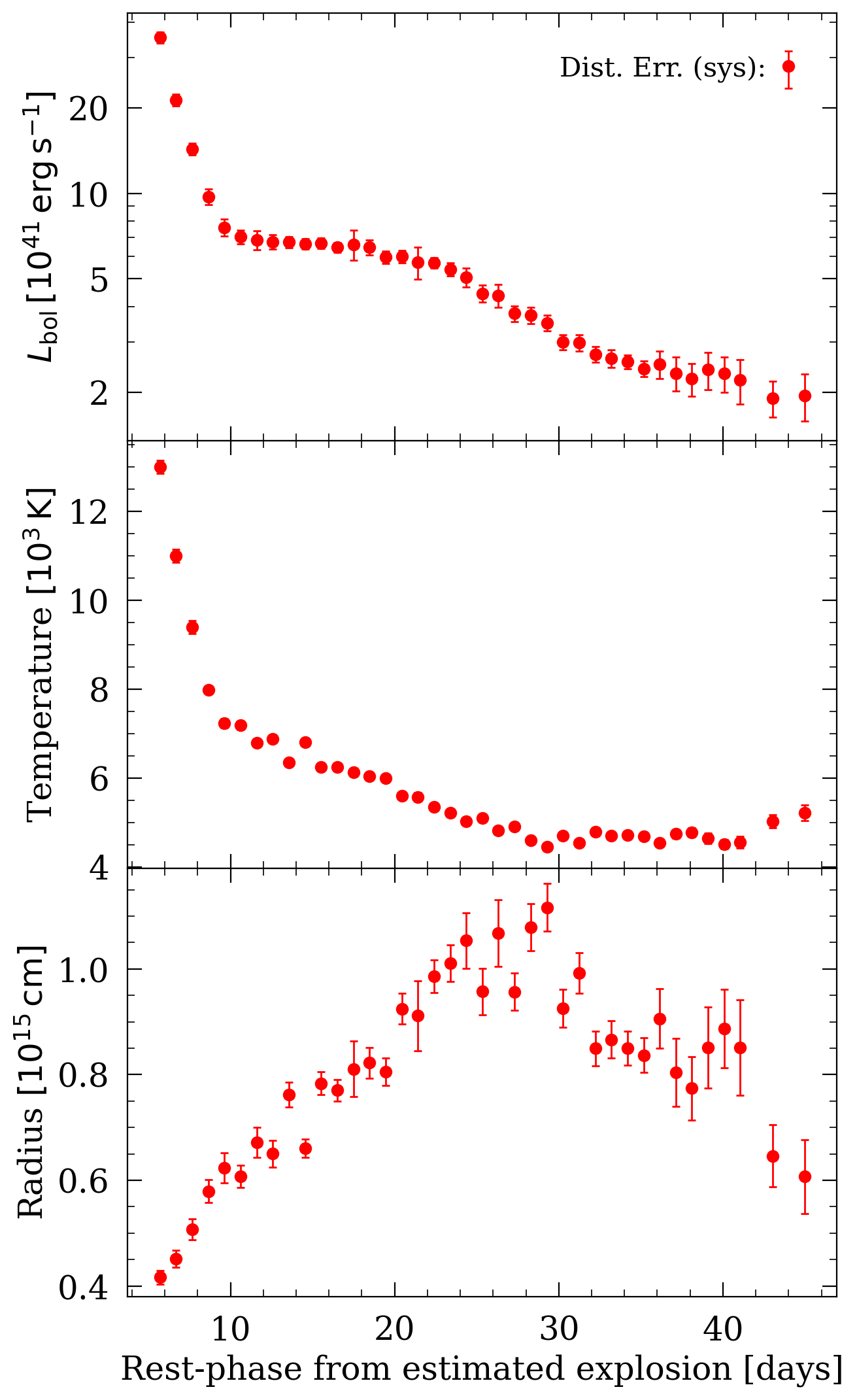}
    \caption{\textit{Top}: Bolometric light curve of SN~2021bxu from integrating the flux in available band with blackbody extrapolation. The systematic error due to the uncertainty in distance is shown as $L^{+13\%}_{-16\%}$. \textit{Middle}: Best-fit temperature from the blackbody fits. \textit{Bottom}: Radius calculated using the best-fit blackbody temperature and bolometric luminosity.}
    \label{fig:L_T_R_BB}
\end{figure}

Our multi-band photometry is used to determine the properties of the SN such as its luminosity, radius and photospheric temperature evolution. We construct the bolometric light curve for SN~2021bxu by using the multi-band photometry in the Swope $BVugri$ bands after converting magnitudes to monochromatic fluxes, correcting for the Milky-Way extinction of $E(B - V)_{\mathrm{MW}} = 0.014\,\mathrm{mag}$, and using the distance of $D_\mathrm{L} = 72 \pm 5 \, \mathrm{Mpc}$ from Table~\ref{tab:21bxu_properties}. We use the magnitude offsets and filter profiles from the CSP webpage\footnote{\url{https://csp.obs.carnegiescience.edu/data/filters}}. If a certain band lacked observations on a given epoch, we interpolated using Gaussian Processes with the \texttt{scikit-learn} \citep{scikit-learn} Python library. We fit each epoch's spectral energy distribution (SED) with a Planck blackbody function. The bolometric luminosity is computed by directly integrating the flux density into the available bands and using the blackbody fits to extrapolate to the unobserved wavelengths. The effective photospheric temperature is the best-fit temperature from the blackbody fits, and the radius is then computed from luminosity and temperature using the Stefan–Boltzmann law. The errors on luminosity, temperature, and radius are derived from a Monte Carlo procedure using the errors of the original photometry. Photometric precision is high with errors $<1\%$ which means majority of the systematic errors come from the uncertainty in the distance estimate and the assumption of a blackbody SED. The uncertainty in the distance corresponds to a fractional uncertainty in luminosity of $L^{+13\%}_{-16\%}$, which would cause the light curve to shift up or down systematically. Figure~\ref{fig:L_T_R_BB} shows the bolometric luminosity, temperature, and radius evolution.

The bolometric luminosity starts at $L_{\mathrm{bol}} \sim 3.5 \times 10^{42}\, \mathrm{erg}\, \mathrm{s^{-1}}$ at discovery, drops down to $L_{\mathrm{bol}} \sim 6.6 \times 10^{41}\, \mathrm{erg}\, \mathrm{s^{-1}}$ during the initial decline, stays flat at that value for $\sim$10\,days defining a plateau, and then declines again. The radius increases almost linearly with time initially until $\sim$25 days past explosion, indicating that the black-body extrapolation is reasonable and the black-body radius roughly follows the photospheric radius. At later times, the black-body radius starts declining, although this may lack physical meaning since the black-body approximation is not good at these times.

The photospheric temperature drops from 13,500\,K to 7000\,K during the initial photometric decline, suggesting a rapid cooling of the ejecta, and steadily declines thereafter. \citet[][]{martinez22} show that SNe~II have a roughly constant temperature evolution during their plateau at $\sim$6000\,K due to H-recombination. The temperature evolution of SN~2021bxu during the plateau is not constant; it decreases from $\sim$7000\,K to $\sim$5000\,K. Combined with the lack of H$\alpha$ emission, this suggests that H-recombination is likely not responsible for the observed plateau in the light curve, unlike SNe~II.

\begin{figure}
	\includegraphics[width=\columnwidth]{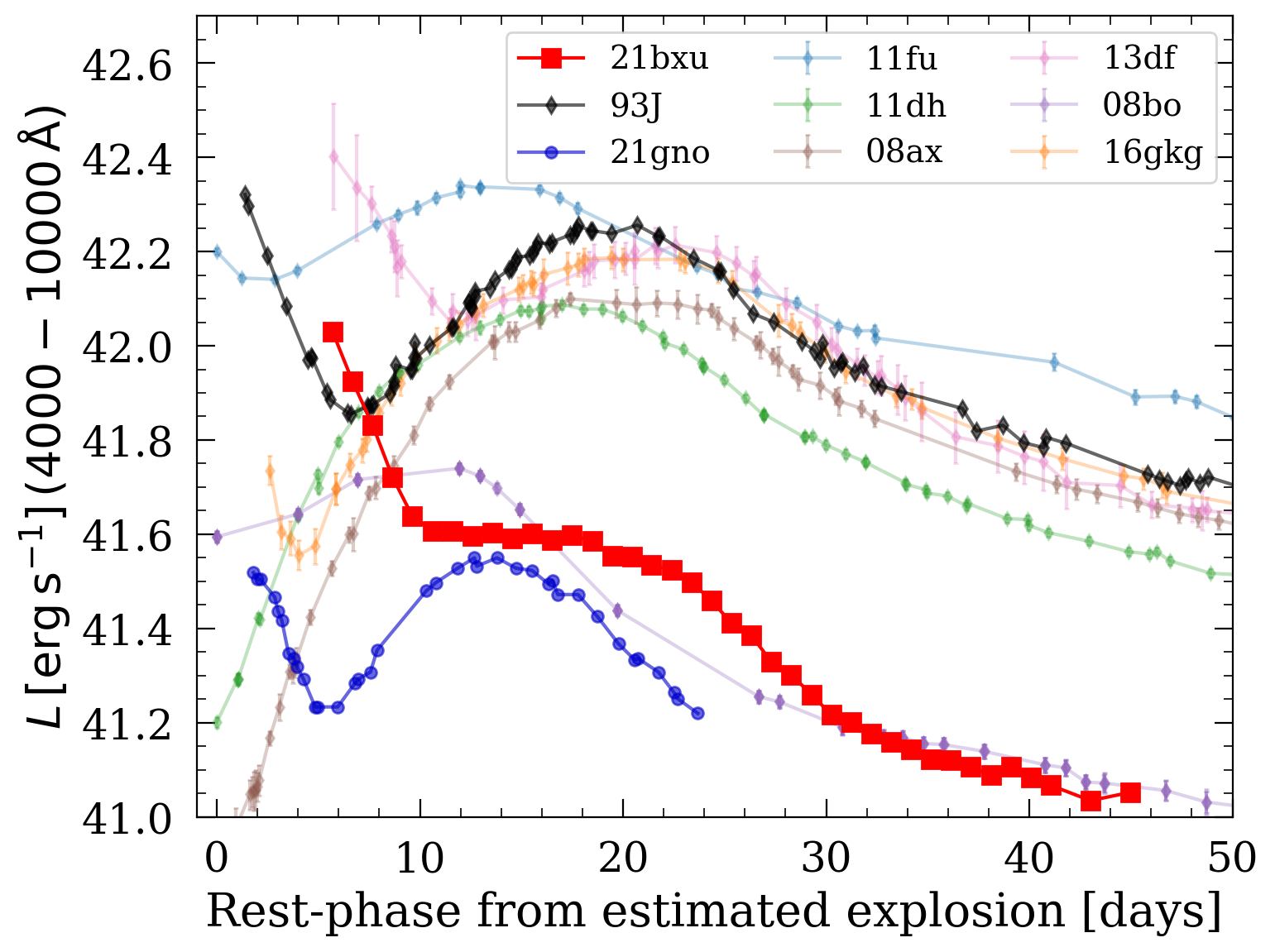}
    \caption{Pseudo-bolometric light curve of SN~2021bxu shown in red compared with a sample of Type IIb SNe from \citet{prentice20} and Ca-rich SN~2021gno from \citet{Ertini23} in the range $4000 - 10000$ \AA. The time axis for all SNe has the zero-point at the estimated time of explosion.}
    \label{fig:prentice_bolo}
\end{figure}

Pseudo-bolometric light curves are used for a direct comparison to similar SNe. Instead of the full wavelength range, pseudo-bolometric light curves are defined only within a finite range usually covering the wavelengths of the observed bandpasses used. We compute the pseudo-bolometric light curve by integrating the fluxes in the range $4000 - 10000$\,\AA\, and compare it with a sample from \citet{prentice20} and SN~2021gno \citep{Ertini23}, due to their potential similarities in the light curve shape, as seen in Figure~\ref{fig:prentice_bolo}. The slope of the initial decline is similar to that of SN~1993J and SN~2021gno; however, they both have a distinct second peak, whereas SN~2021bxu shows a plateau. SN~2021gno has the lowest luminosity with rapid photometric evolution and low explosion energy, $^{56}$Ni mass and ejecta mass, characteristic of Ca-rich SNe \citep{Ertini23}. We note again that SN~2021bxu is unique due to its low peak luminosity at $\log(L_{\mathrm{pseudo}}/\mathrm{erg\,s^{-1}}) = 42.0$ and a distinct plateau phase at $\log(L_{\mathrm{pseudo}}/\mathrm{erg\,s^{-1}}) \sim 41.6$ from $\sim$10 to $\sim$20 days post-explosion. This plateau is possibly due to an underlying secondary peak from the radioactive decay of $^{56}$Ni similar to SN~1993J and SN~2021gno (further discussed in Sections \ref{sec:modelling} and \ref{sec:discussion}).

\section{Spectroscopic Analysis} \label{sec:spec_anal}

\subsection{Line IDs}
Due to the homologous expansion of the ejecta, measuring line velocities as a function of time allows us to examine the chemical composition of the ejecta, and understand the structure and the mixing within the explosion. We identify the lines by comparing the spectral features to literature and cross-checking with measured velocities (Section~\ref{subsec:line_vels}). There is a total of ten optical spectra ranging from 7 to 31 days after estimated time from explosion. This allows us to explore the velocity evolution of the absorption features. 

The main absorption features are labeled in the spectra shown in Figure~\ref{fig:21bxu_spectra}.  
We identify strong absorption features from \HeI\,$\lambda 5876$ and \HeI\,$\lambda 6678$ along with weaker hydrogen Balmer series (H$\alpha$\,$\lambda 6563$, H$\beta$\,$\lambda 4861$, and H$\gamma$\,$\lambda 4340$).
The presence of strong helium with some hydrogen is characteristic of a Type IIb SN and confirms the typing of SN~2021bxu as Type IIb. Absorption features from heavier elements such as \OI\,$\lambda 7774$, \SiII\,$\lambda\lambda4130, 5972, 6355$, \CaII\,H\&K\,$\lambda\lambda3934, 3969$, IR-triplet\,$\lambda\lambda8498, 8542, 8662$, and a forest of \FeII\, lines including the strong \FeII\,$\lambda5169$, are also present. We also identify absorption features from neutron capture elements such as \ScII\,$\lambda\lambda5527, 5698, 6280$ and \BaII\,$\lambda6142$ in SN~2021bxu. Disentangling whether these are r- or s-process elements is beyond the scope of this study. Interestingly, the neutron capture elements are usually found in SN~1987A-like objects \citep[][]{Williams87,tsujimoto01}. Sc and Ba have also been observed in the subluminous SN Ia PTF~09dav \citep{Sullivan11}. This demonstrates that, although SN~2021bxu is formally classified as a SN~IIb, it has some spectroscopic similarities to SNe~II and SNe~Ia. 

\subsection{Line Velocities} \label{subsec:line_vels}
We measure the Doppler shift of the minima of the spectral features to determine the velocities and chemical composition of the ejecta. The velocities for \HeI\,$\lambda\lambda 5876, 6678, 7065$, H$\alpha$ $\lambda 6563$, \FeII\,$\lambda 5169$, and \OI\,$\lambda 7774$ are computed using \texttt{misfits}\footnote{\url{https://github.com/sholmbo/misfits}} \citep{Holmbo2020}, which is an interactive tool used to measure spectral features in spectra of transients and calculate their errors. Within \texttt{misfits}, we smooth the spectrum by applying a low-pass filter to the Fourier-transformed data as described in \citet[][]{marion09} and obtain best-fit Gaussians to the absorption features with a fixed local continuum. The best-fit mean of the Gaussian with the associated error from Monte Carlo iterations is taken to be the absorption feature's observed wavelength which is then converted to a line velocity. 

\begin{figure}
	\includegraphics[width=\columnwidth]{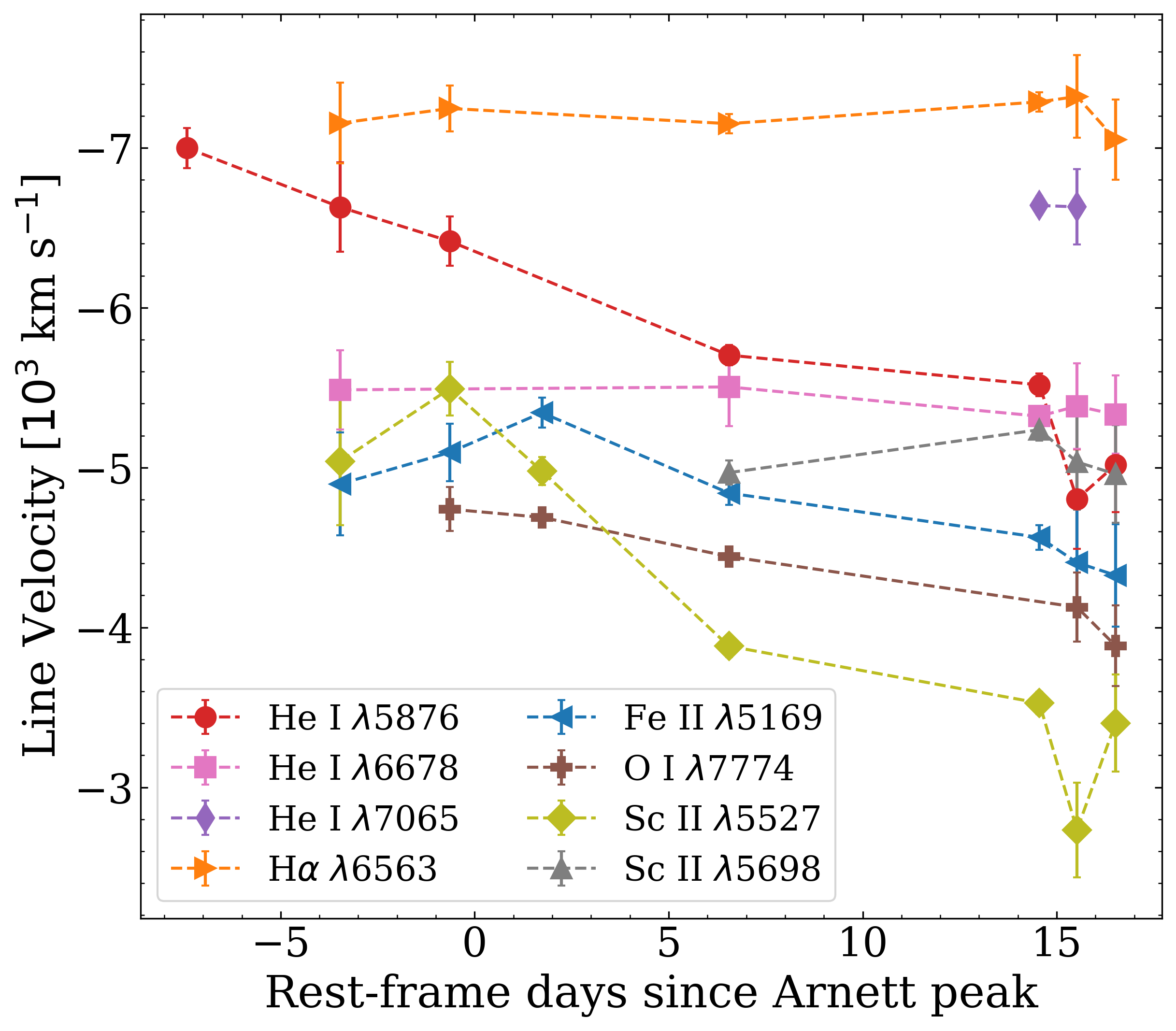}
    \caption{Evolution of line velocities of \HeI\,$\lambda 5876$,\,$\lambda 6678$,\,$\lambda 7065$, H$\alpha$ $\lambda 6563$, \FeII\,$\lambda 5169$, \OI\,$\lambda 7774$, \ScII\,$\lambda 5527$, and \ScII\,$\lambda 5698$ measured from the spectra using \texttt{misfits}. Only those epochs are included for each line where a clear measurement was possible. Note that the \FeII\,$\lambda 5169$ velocity evolution may be uncertain because it comes from a blend of spectral features in the forest of Fe lines.}
    \label{fig:line_vels}
\end{figure}

Figure~\ref{fig:line_vels} shows the velocity evolution of the selected features. The velocity of the H$\alpha$\,$\lambda 6563$ feature stays constant at $\sim 7200\, \mathrm{km\,s^{-1}}$ from $\sim$10 to $\sim$30 days after explosion, demonstrating that at these early phases the photosphere has already reached the bottom of the H layer. Interestingly, the feature from \SiII\,$\lambda6355$ as seen in Figure~\ref{fig:21bxu_spectra} could have some contribution from a high velocity H$\alpha$\,$\lambda 6563$ component at $\sim13{,}000\, \mathrm{km}\,\mathrm{s^{-1}}$. This also matches with the small feature just bluer to H$\beta$\,$\lambda 4861$. Therefore, there could be a detached high velocity H component in the ejecta. It may be the case that this comes from interaction with an extended envelope. 

\begin{figure*}
	\includegraphics[width=\textwidth]{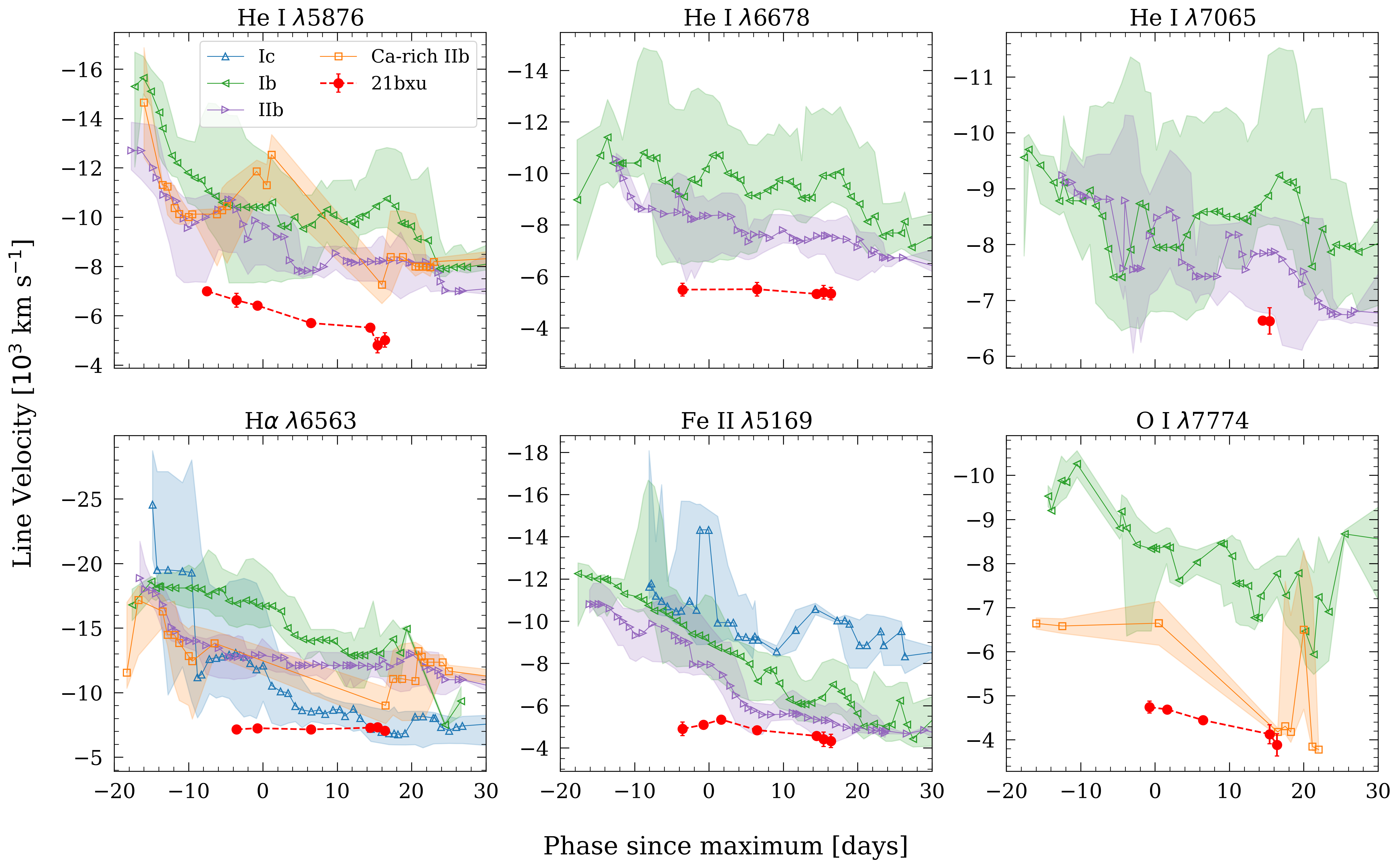}
    \caption{Evolution of line velocities of SN~2021bxu compared to a sample of SNe~IIb, Ib, Ic from \citet[][]{liu16} and Ca-rich IIb SNe from \citet[][]{das22}. The open markers show a rolling median for each SN type with a bin size of five days where the shaded regions represent the $16^{\mathrm{th}}$ and $84^{\mathrm{th}}$ percentiles of the distribution in the bin indicating the dispersion of velocities. In the case of double-peaked light curves, the phases are relative to the second bolometric peak. Values for SN~2021bxu are shown as filled red circles.}
    \label{fig:line_vels_comp}
\end{figure*}

The velocity of the \HeI\,$\lambda 5876$ line decreases as a function of time as the photosphere recedes, from $\sim7000\,\mathrm{km\,s^{-1}}$ at day $+7$ to $\sim5700\,\mathrm{km\,s^{-1}}$ at day $+21$. At these phases the \HeI\,$\lambda 5876$ line velocity follows the photospheric velocity, after which it plateaus, demonstrating that the base of the He layer is at $\sim5500\,\mathrm{km\,s^{-1}}$. This \HeI\,$\lambda 5876$ line is used to break the Arnett degeneracy in modelling the bolometric light curve (see Section~\ref{sec:modelling}). The velocity of the \HeI\,$\lambda 7065$ line is only measured in the later two spectra, thus not showing its early evolution and making it unreliable for analysis. In SE-SNe the He lines require non-thermal excitation. Therefore they get stronger over time as the density of the ejecta decreases and the mean free path of the $\gamma$-rays can increase \citep{lucy91}. 

The \OI\,$\lambda 7774$ line shows the lowest velocity at $< 4700\, \mathrm{km\,s^{-1}}$ throughout the time range. This is expected as the progenitor star would have a layered structure prior to explosion where heavier elements are further in, towards the center of the star. 
The \FeII\,$\lambda 5169$ feature shows a similar decline in velocity to that of other lines at later times. However, it has slightly higher velocities than oxygen, possibly caused by the fact that there could be primordial Fe-mixing throughout the progenitor star, leading to Fe in the top layers. Due to the Einstein coefficient values of \FeII\,$\lambda 5169$, only a small abundance is sufficient to produce the opacity needed for a strong line. 
The \ScII\,$\lambda 5698$ line shows higher velocities than the \ScII\,$\lambda 5527$ and \FeII\,$\lambda 5169$ lines. This is potentially due to a metallicity effect or mixing phenomena in the ejecta. However, further spectral modelling is required to disentangle this.

In Figure~\ref{fig:line_vels_comp}, we compare our measurements of line velocities with \citet[][]{liu16}, which provides a spectroscopic sample of SE-SNe, and with \citet[][]{das22}, which provides spectroscopic measurements for Ca-rich~IIb SNe. We compute a rolling median for each SN type with a bin size of five days where the shaded regions represent the $16^{\mathrm{th}}$ and $84^{\mathrm{th}}$ percentiles of the distribution in the bin indicating the dispersion of velocities. The sample of SNe~IIb has velocities lower than those of SNe~Ib for most of the lines with Ca-rich~IIb SNe showing velocities similar to those of SNe~IIb. All line velocities for SN~2021bxu are consistently lower than the median of all types by at least $\sim1500\,\mathrm{km\,s^{-1}}$, emphasizing its uniqueness and implying that the kinetic energy of SN~2021bxu is lower than that of typical SE-SNe.

\subsection{Time of Explosion} \label{subsec:t_exp}
Using line velocities of \HeI\,$\lambda 5876$ from the spectra (see Section \ref{subsec:line_vels}) along with radii computed from the black-body fits, we trace back to the time of explosion. Assuming a homologous expansion of ejecta, the relation is given by $t_{\mathrm{exp}} \propto R_{\mathrm{BB}}/v_{\mathrm{ph}}$, which becomes $t_{\mathrm{exp}} \approx R_{\mathrm{BB}}/v_{\mathrm{ph}}$ for small initial radius compared to the post-explosion radius of the ejecta. We expect the estimated explosion time to fall between the last non-detection (6.01 days before discovery) and the discovery. We choose measurements from two spectra where the \HeI\,$\lambda 5876$ velocity is still linearly declining and has not plateaued, ensuring that these values are representative of the photosphere and not the base of the He layer. Using the values $v_{\mathrm{ph}} = \{6600 \pm 300, 6400\pm 200\}\, \mathrm{km\,s^{-1}}$ and $R_{\mathrm{BB}} = \{(6.1 \pm 0.2) \times 10^{14}, (7.6 \pm 0.3) \times 10^{14} \}\, \mathrm{cm}$, we obtain the averaged time of explosion within the expected window, at $5.0\, \pm 0.4\, \mathrm{days}$ before discovery, on $\mathrm{MJD}\, 59246.3\, \pm 0.4\, \mathrm{days}$. This is the estimated time of explosion we adopt throughout this paper. The uncertainty on the time of explosion is purely statistical and is appropriately propagated from the uncertainties on radius and velocity. 

\subsection{Spectral Comparison with Similar Supernovae}
\begin{figure*}
	\includegraphics[width=\textwidth]{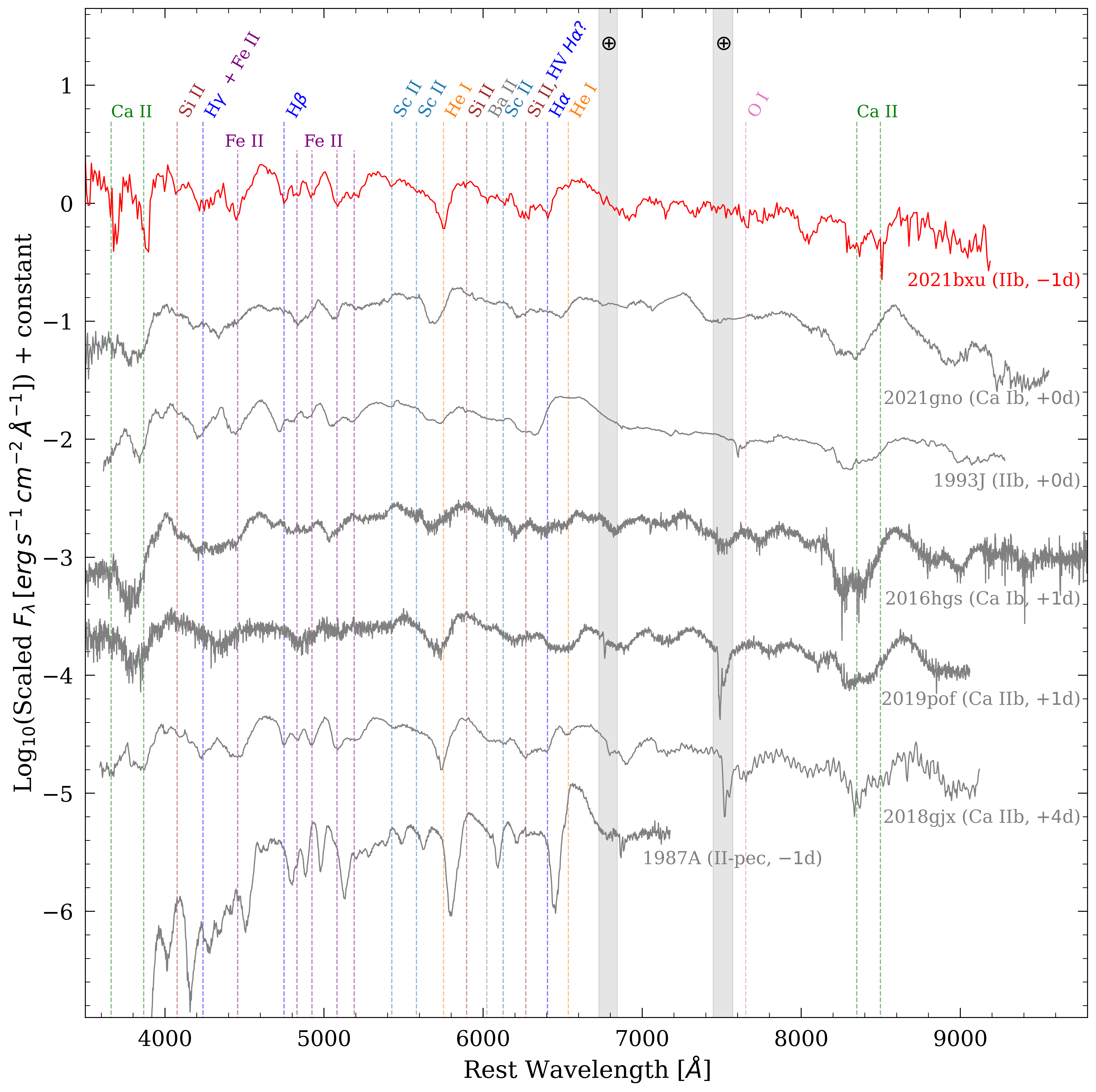}
    \caption{NOT spectrum of SN~2021bxu near the $^{56}$Ni peak shown in red compared to the NOT spectrum of Ca-rich Ib SN~2021gno \citep{Ertini23}, the Asiago Observatory spectrum of Type IIb SN~1993J \citep[][]{Barbon95}, the Keck-LRIS spectrum of Ca-rich Type Ib SN~2016hgs \citep[][]{de18}, the Palomar-200inch spectrum of Ca-rich Type IIb SN~2019pof \citep{das22}, the NTT spectrum of Ca-rich Type IIb SN~2018gjx \citep[][]{prentice20}, and the International Ultraviolet Explorer (IUE) spectrum of Type II SN~1987A \citep[][]{pun95}. Major absorption features in SN~2021bxu are marked on the plot along with the two telluric regions shown in gray bands. All phases in this figure are relative to the $^{56}$Ni peak of each SN.}
    \label{fig:spec_comp}
\end{figure*}

\begin{figure}
	\includegraphics[width=\columnwidth]{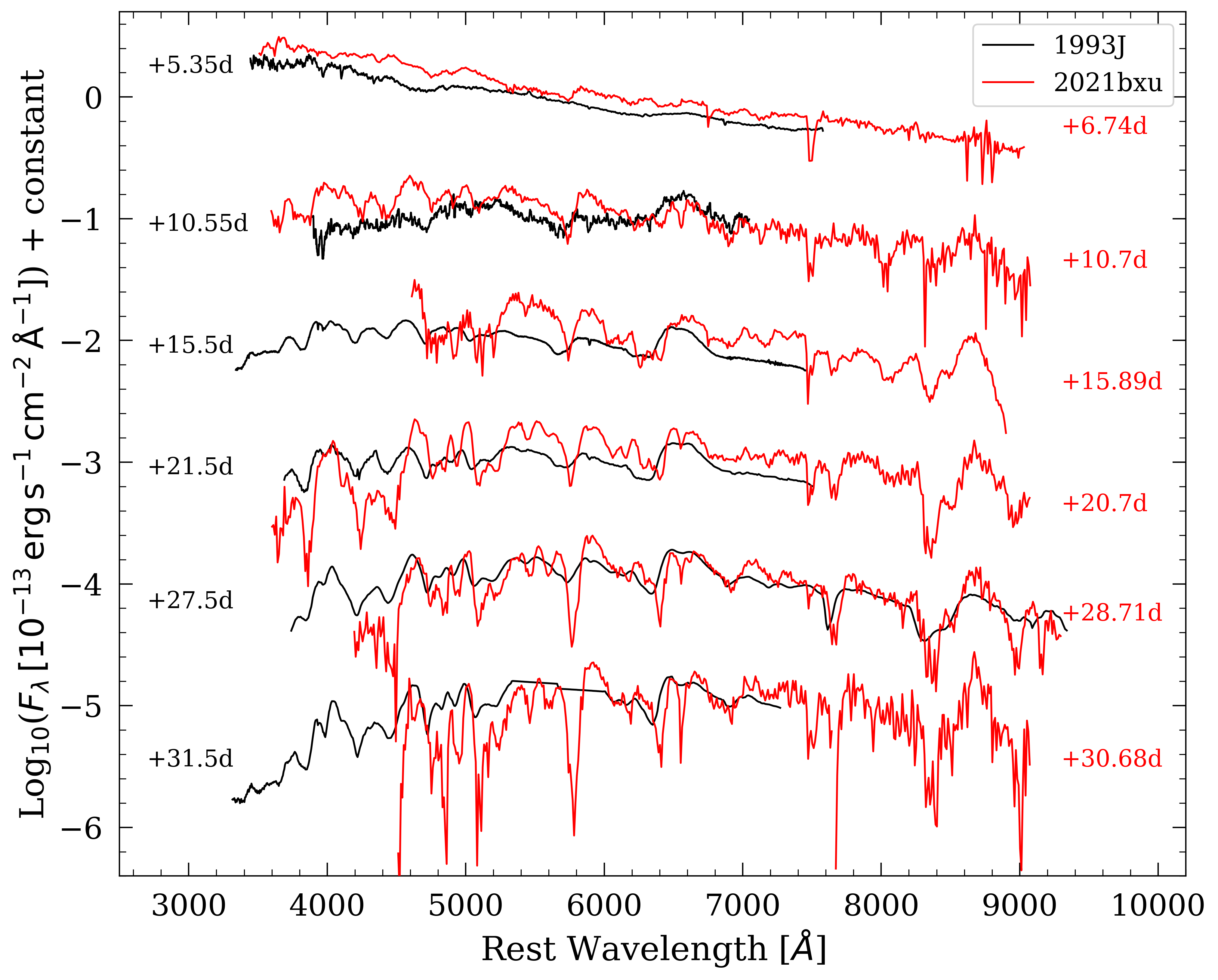}
    \caption{Spectral comparison of SN~2021bxu with SN~1993J. The spectra for SN~1993J are shown in black with the epochs labeled on the left. The spectra for SN~2021bxu are shown in red with the epochs labeled on the right. For both objects, the epochs are relative to the estimated time of explosion. The spectra at $+5.35$, $+10.55$, $+21.5$, and $+27.5$ days for SN~1993J are from \citet[][]{Barbon95}, $+15.5$ days is from \citet[][]{matheson00b}, and $+31.5$ days is from \citet[][]{matheson00a}.}
    \label{fig:93J_spec_comp}
\end{figure}

In this section, we compare the spectrum of SN~2021bxu near the peak due to $^{56}$Ni with similar SNe near their peaks. Figure~\ref{fig:spec_comp}, along with the line IDs, shows spectra for a Type II-pec SN~1987A, a Type IIb SN~1993J, a Ca-rich Type Ib SN~2016hgs and SN~2021gno, and Ca-rich Type IIb SNe 2018gjx, and 2019pof. These SNe, except SN~1987A, are chosen for comparison because of their similarities either in light curve shape or spectral features. SN~1987A is included for its strong H lines showing recombination.

SN~2021bxu is dissimilar to SN~1987A due to the lack of strong hydrogen Balmer features (H$\alpha$\,$\lambda 6563$, H$\beta$\,$\lambda 4861$, and H$\gamma$\,$\lambda 4340$) and also due to the absence of the \NaI\,D\,$\lambda\lambda5890, 5896$ doublet lines. However, SN~2021bxu does show features from the neutron capture elements \ScII\,$\lambda\lambda5527, 5698, 6280$ and \BaII\,$\lambda6142$ which are typically seen in SN~1987A-like objects. There are also no signatures of strong H emission, which is usually seen in SNe~IIP, hence providing further evidence that the plateau in SN~2021bxu is unlikely to be caused by H-recombination. SN~2021bxu is more similar to SN~1993J, reinforcing the typing of SN~2021bxu as a Type IIb. 

Comparing the spectral time-series of SN~2021bxu with SN~1993J at similar epochs in Figure~\ref{fig:93J_spec_comp}, we note that both SNe display similar absorption lines in their spectra. SN~1993J shows broader features at higher velocities between $10{,}000$ and $16{,}000\,\mathrm{km\,s^{-1}}$ \citep[][]{garnavich94} compared to the velocities of SN~2021bxu that we measure in the range $4000 < v < 7000\,\mathrm{km\,s^{-1}}$. This points to SN~2021bxu having lower energies and masses than SN~1993J. Although they both show similar features, SN~2021bxu evolves more quickly showing strong metal features from \CaII\,[IR-triplet] and \OI\,$\lambda7774$ by day 30. The \HeI\,$\lambda 5876$ and \HeI\,$\lambda 6678$ features also quickly get deeper, showcasing the fast evolution of SN~2021bxu. Moreover, SN~1993J shows weaker \HeI\,$\lambda 5876$ absorption than SN~2021bxu but H$\alpha$\,$\lambda 6563$ absorption of similar depth. Due to their spectral similarities, it may be the case that SN~2021bxu and SN~1993J are similar objects, both with a large initial decline but with different amount of $^{56}$Ni, energy and mass. We discuss this further in Section~\ref{sec:discussion}.

The Ca-rich SNe show characteristically strong \CaII\,compared to \OI. During the photometric phases, SN~2021bxu shows comparably strong \CaII\,absorption to that of the Ca-rich Ib and IIb SNe. One of the defining properties of Ca-rich transients is that they quickly transition to nebular phase marked by \CaII\,and \OI\,emission. For instance, SN~2019ehk (Ca-rich IIb) started exhibiting nebular features as early as 30 days after explosion. SN~2021bxu does not show any nebular features within $\sim$30 days after explosion. Due to the lack of late phase spectra for SN~2021bxu, we cannot directly compare it to Ca-rich SNe using the [\CaII]/[\OI] ratio. However, the hydrogen-rich SN~2021bxu shows dissimilarities to the Ca-rich Ib SNe (SN~2021gno, SN~2016hgs), which lack hydrogen in their spectra, but is spectroscopically similar to IIb (SN~1993J) and hydrogen- and Ca-rich IIb SNe (SN~2018gjx) near peak.

\section{Modelling}
\label{sec:modelling}

\begin{figure}
	\includegraphics[width=\columnwidth]{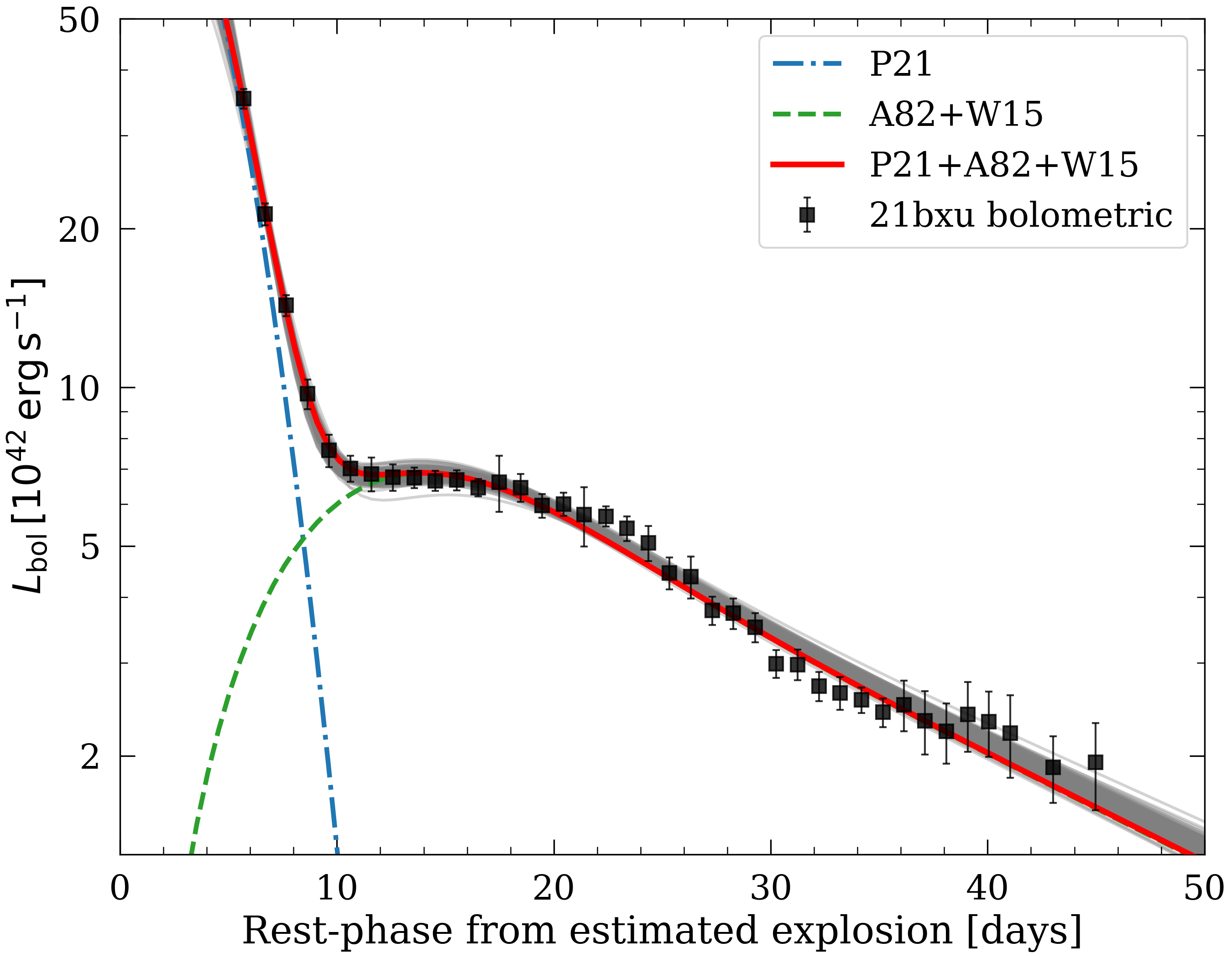}
    \caption{Bolometric light curve of SN~2021bxu shown in black fit using a two-component model: shock cooling and $^{56}$Ni decay including $\gamma$-ray leakage. The initial decline is fit with the shock interaction with extended envelope model from \citet[][]{piro21} shown as the dotted-dashed line and $^{56}$Ni decay comes from the model from \citet[][]{arnett82} following an additional correction for $\gamma$-ray leakage as shown by \citet[][]{wheeler15} shown as the dashed line. The best-fit model is the solid red line and the gray lines show 100 randomly selected parameters from the Monte Carlo samples signifying the uncertainty on the best-fit.}
    \label{fig:bolo_fits}
\end{figure}

To understand the origin of the unique shape of SN~2021bxu's light curve, we analyze the explosion by fitting the bolometric and pseudo-bolometric light curves of SN~2021bxu with SN explosion models from the literature. We do this by using a two-component model where the first component is the initial cooling phase, and the second component is the radioactive decay of $^{56}$Ni including $\gamma$-ray leakage. 

The analytic model we consider for the initial cooling phase is from \citet[hereafter P21]{piro21}. \citetalias{piro21} describes the shock interaction with the extended material surrounding the progenitor star once it is in thermal equilibrium and in homologous expansion phase, given a two-component density profile with steep radial dependence in the outer region and shallower radial dependence in the inner region. The free parameters in this model are the mass and radius of the extended material ($M_e$ and $R_e$, respectively) with $E_e$, the energy imparted by the SN shock to the extended material, depending on the total explosion energy, ejecta mass, and mass of the extended material. The analytic model from \citetalias{piro21} is an improvement over previous similar models \citep[e.g.,][]{Piro13,piro15} as it better matches the observations in the early shock-cooling emission and is tested against numerical models.

Together with the \citetalias{piro21} model, we use the analytic models from \citet[hereafter A82]{arnett82} for the plateau/secondary peak from $^{56}$Ni decay, which provides an estimate of the total ejecta mass $\left( M_{\mathrm{ej}}\right)$, $^{56}$Ni mass $\left( M_{\mathrm{Ni}}\right)$, and total explosion energy of the SN $\left( K_{\mathrm{ej}}\right)$. We adopt a mean optical opacity of $\bar \kappa_{\mathrm{opt}} = 0.1\,\mathrm{cm^2\,g^{-1}}$ corresponding to electron scattering. The \citetalias{arnett82} model fits for a degenerate parameter that depends on $M_{\mathrm{ej}}$ and $K_{\mathrm{ej}}$ as $ M_{\mathrm{ej}}^{3/4} / K_{\mathrm{ej}}^{1/4}$. This degeneracy is broken by using the velocity of the \HeI\,$\lambda 5876$ line near maximum of the $^{56}$Ni peak to obtain the ejecta expansion velocity \citep[see Section~5.3 of][]{Dessart16}. After interpolating between the spectral epochs, we use $v_{\mathrm{ej}} = 4900 \pm 200\, \mathrm{km}\, \mathrm{s^{-1}}$ near peak. We include an additional correction for $\gamma$-ray leakage at late times as shown by \citet[hereafter W15]{wheeler15} with a multiplicative factor of $1 - e^{-(T_0/t)^2}$ where $T_0$ is the characteristic time-scale for the $\gamma$-ray leakage, depending on $M_{\mathrm{ej}}$ and $K_{\mathrm{ej}}$ as
\begin{equation}
    T_0 = \left( \frac{C \kappa_{\gamma} M_{\mathrm{ej}}^2}{K_{\mathrm{ej}}} \right)^{1/2} ,
\end{equation}
where $C$ is a dimensionless structure constant dependent on the slope of the density profile (typically $C\sim0.05$) and $\kappa_{\gamma}$ is the opacity to $\gamma$-rays (fiducial value of $\kappa_{\gamma} = 0.03\,\mathrm{cm^2\,g^{-1}}$).

We adopt a chi-squared minimization approach to fit the bolometric and pseudo-bolometric light curves of SN~2021bxu and obtain the best-fit parameters. \citet{Nakar14} give a relation between the \citetalias{piro21} parameters $E_e$ and $M_e$ along with a dependency on the \citetalias{arnett82} parameters $M_\mathrm{ej}$ and $K_{\mathrm{ej}}$,
\begin{equation}
    E_e \approx 2 \times 10^{49} \left(\frac{K_{\mathrm{ej}}}{10^{51}\,\mathrm{erg}}\right) \left(\frac{M_{\mathrm{ej}}}{3\,M_{\odot}}\right)^{-0.7} \left(\frac{M_e}{0.01\,M_{\odot}}\right)^{0.7}\, \mathrm{erg}.
    \label{eq:E_e}
\end{equation}
Therefore, we eliminate $E_e$ from the fitting routine and later solve for it using Eq.~\ref{eq:E_e}. With only $\sim$5 data points in the initial decline, the parameters from \citetalias{piro21} for the early light curve are difficult to constrain. However, the model fits well the later part where \citetalias{arnett82}+\citetalias{wheeler15} dominates and there are more points to fit. Changing the initial decline of the light curve barely affects the second component assumed from $^{56}$Ni decay. The best-fit parameters are given by the maximum-likelihood values and the uncertainties on best-fit parameters are given by the $16^{\mathrm{th}}$ and $84^{\mathrm{th}}$ percentiles from fitting the model to Monte Carlo resamples of the light curve. The uncertainties for the parameters derived using the best-fit values are propagated appropriately from the uncertainties on the best-fit values. The Monte Carlo resampling ensures that statistical uncertainties from the photometry as well as systematic uncertainties from distance measurement are appropriately considered. 

When fitting the bolometric and pseudo-bolometric light curves, we find that the \citetalias{piro21}+\citetalias{arnett82}+\citetalias{wheeler15} model can successfully describe the data. Figure~\ref{fig:bolo_fits} shows the two components separately as well as the combined fit in red for the bolometric light curve. The surrounding gray region is randomly drawn Monte Carlo samples showing the uncertainty on the best-fit model. The best-fit parameters for the bolometric and pseudo-bolometric light curve fits are listed in Table~\ref{tab:fit_params}. The parameters for the pseudo-bolometric light curve are provided because they are useful in direct comparison with pseudo-bolometric light curves from literature. It should be noted that the pseudo-bolometric light curves do not encompass the total flux and hence one must be cautious when inferring physical parameters from them.

\begin{table}
	\centering
    \caption{Best-fit parameters after fitting the \citetalias{piro21}+\citetalias{arnett82}+\citetalias{wheeler15} model to the bolometric and pseudo-bolometric light curves}
	\label{tab:fit_params}
	\renewcommand{\arraystretch}{1.6}
	\begin{tabular}{lcc}
		\hline
		\hline
		Parameter & Bolometric & Pseudo-bolometric \\
		& & ($4000 - 10000$\,\AA) \\
		\hline
		$M_{\mathrm{Ni}}$ $[\mathrm{M_{\odot}}]$     & $0.029^{+0.004}_{-0.005}$     & $0.017^{+0.002}_{-0.003}$ \\
		$M_{\mathrm{ej}}$ $[\mathrm{M_{\odot}}]$     & $0.61^{+0.06}_{-0.05}$        & $0.55^{+0.06}_{-0.05}$    \\
		$K_{\mathrm{ej}}$ $[10^{49}\, \mathrm{erg}]$ & $8.8^{+1.1}_{-1.0}$           & $7.9^{+1.1}_{-1.0}$       \\
		$E_e$ $[10^{49}\, \mathrm{erg}]$             & $2.0^{+0.4}_{-0.3}$           & $2.1^{+0.6}_{-0.3}$       \\
		$M_e$ $[\mathrm{M_{\odot}}]$                 & $0.065^{+0.005}_{-0.005}$     & $0.076^{+0.011}_{-0.010}$ \\
		$R_e$ $[\mathrm{R_{\odot}}]$                 & $1400^{+300}_{-200}$          & $370^{+100}_{-50}$         \\
		\hline
	\end{tabular}\\
	\begin{flushleft}
	\end{flushleft}
\end{table}

\section{Discussion}
\label{sec:discussion}
In this section we attempt to put our results in context relative to other SN explosions and models. The best-fit parameters for the bolometric light curve are $M_{\mathrm{Ni}} = 0.029^{+0.004}_{-0.005}\,\mathrm{M_{\odot}}$, $M_{\mathrm{ej}} = 0.61^{+0.06}_{-0.05}\,\mathrm{M_{\odot}}$, $K_{\mathrm{ej}} = 8.8^{+1.1}_{-1.0} \times 10^{49}\, \mathrm{erg}$, $E_e = 2.0^{+0.4}_{-0.3} \times 10^{49}\, \mathrm{erg}$, $M_e = 0.065^{+0.005}_{-0.005}\,\mathrm{M_{\odot}}$, and $R_e = 1400^{+300}_{-200}\,\mathrm{R_{\odot}}$ (listed in Table~\ref{tab:fit_params}). SN~2021bxu is a Type IIb SN with low $M_{\mathrm{Ni}}$, low luminosity, and low explosion energy. By comparison with known classes of SNe and models explaining the observed features of SN~2021bxu, we can infer the details of the explosion and the progenitor system.

\begin{figure}
	\includegraphics[width=\columnwidth]{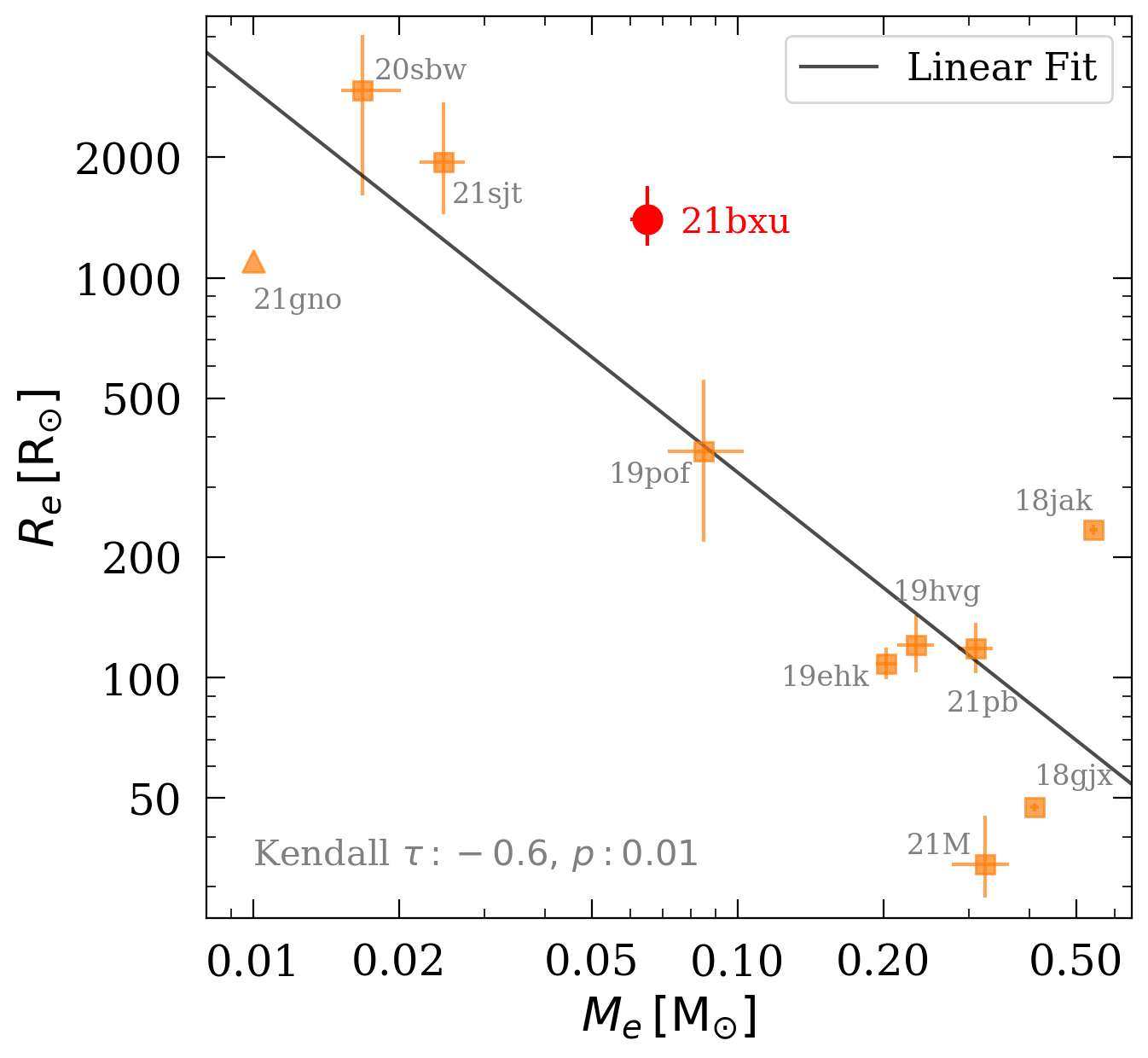}
    \caption{Mass and radius of the extended material ($M_e$ and $R_e$, respectively) from fitting the initial decline of SN~2021bxu compared with the sample of Ca-rich SNe~IIb \citep{das22} and the Ca-rich Ib SN~2021gno \citep{Ertini23}. The values for Ca-rich SNe~IIb (orange squares) and SN~2021bxu (red circle) are obtained from fitting the \citetalias{piro21} model and the values for SN~2021gno (orange triangle) are obtained using a hydrodynamic code from \citet{Bersten11}. We note a trend of decreasing radius with increasing mass with a linear best-fit shown as the black line and Kendall $\tau$ correlation statistic given.}
    \label{fig:Me_Re_comp}
\end{figure}

The fit to the \citetalias{piro21} model shows that the extended material surrounding the progenitor of SN~2021bxu had a large radius ($R_e$) and low mass ($M_e$). We compare this to the best-fit masses and radii of the Ca-rich IIb sample from \citet{das22}, who use the same model for the initial decline. Figure~\ref{fig:Me_Re_comp} shows SN~2021bxu along with the sample of Ca-rich IIb SNe. We see a clear trend of decreasing radius with increasing mass. The Ca-rich Ib SN~2021gno also falls along this trend with $R_e = 1100\,\mathrm{R_{\odot}}$ and $M_e = 0.01\,\mathrm{M_{\odot}}$. For a simple check, we show a linear best-fit and a negative correlation using the Kendall $\tau$ test. The best-fit line is given by $\log_{10}(R_e) = -0.96 \log_{10}(M_e) + 1.55$ with a root-mean-square deviation of $0.16\,\mathrm{dex}$. The Kendall $\tau$ test gives $\tau = -0.6$ with a $p$-value of 0.01. Modelling and possible physical origins of this correlation will be the subject of a future work.

Theory suggests that a plateau can arise in the light curve of a SE-SN $\sim$1 day after shock-breakout when the cooling of the photosphere slows down, allowing the recombination of ejecta layers, primarily He \citep[][]{dessart11}. Observed in simulations by \citet[][]{dessart11}, this plateau is found at $\log (L_{\mathrm{bol}}/\mathrm{erg\,s^{-1}}) \sim 41$ and lasts for $\sim$10 days until the SN either re-brightens for $^{56}$Ni-rich SNe or fades away for $^{56}$Ni-poor SNe. The plateau in SN~2021bxu is observed for a time-scale comparable to that of He-recombination, however, the timing of occurrence differs. The plateau caused by He-recombination occurs soon after explosion, whereas, the plateau in SN~2021bxu's light curve is not apparent until $\sim$10 days after explosion. Moreover, the luminosity of SN~2021bxu at the plateau is $\log (L_{\mathrm{bol}}/\mathrm{erg\,s^{-1}}) \sim 41.8$, almost an order of magnitude higher. This suggests that the observed plateau in SN~2021bxu is likely not due to He-recombination.

SN~2021bxu shows photometric and spectroscopic similarities to SN~1993J but with lower total mass and explosion energy. SN~1993J shows an initial decline due to the post shock cooling through a thin H-rich envelope of extended material and the main peak from $^{56}$Ni decay. \citet[][]{bersten12} and \citet[][]{prentice16} estimate $M_{\mathrm{Ni}}$ in the range $0.084 - 0.15\,\mathrm{M_{\odot}}$ and $M_{\mathrm{ej}} = 3.3\,\mathrm{M_{\odot}}$ from fitting the bolometric light curve of SN~1993J. This is $\sim3 - 5$ times more $^{56}$Ni mass than in SN~2021bxu and $\sim$6 times more ejecta mass. Higher $M_{\mathrm{Ni}}$ and $M_{\mathrm{ej}}$ lead to the main $^{56}$Ni powered light curve to be more luminous and broader, respectively, which is seen in the light curve of SN~1993J. The light curve of SN~1993J has been described by a binary model with both stars having a mass of $\sim15\,\mathrm{M_{\odot}}$ \citep[][]{podsiadlowski93,nomoto93}. The authors conclude that SN~1993J had a G8-K0 yellow supergiant progenitor with a binary companion. The strong initial decline can be explained by the explosion of this progenitor that has experienced mass-loss due to winds, or, more likely, through mass transfer to a companion star. Given that SN~2021bxu is a similar object to SN~1993J at least in the initial decline and a two-component light curve, SN~2021bxu may be conspiring to show a plateau instead of an evident second peak owing to the small amount of $^{56}$Ni produced and the overall lower energies.

\begin{figure}
	\includegraphics[width=\columnwidth]{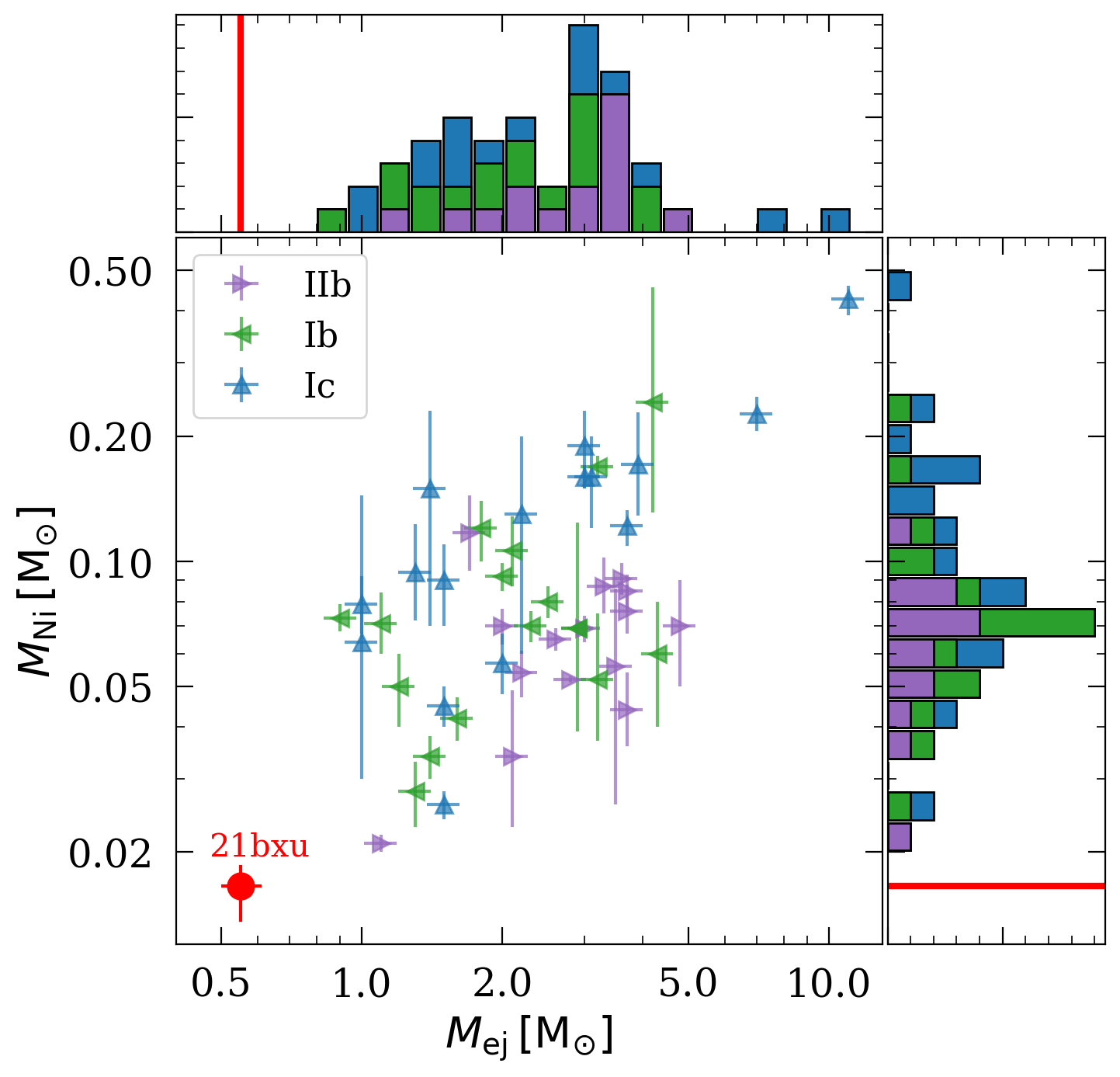}
    \caption{Distributions of $M_{\mathrm{Ni}}$ and $M_{\mathrm{ej}}$ derived using the pseudo-bolometric light curves in the range  $4000 - 10000$\,\AA\, for the sample of SNe~IIb, Ib, Ic from \citet[][]{prentice16,prentice19}. The values for SN~2021bxu are shown in red.}
    \label{fig:P16-19_pseudo_comp}
\end{figure}

For a direct comparison with other studies without dealing with the SED flux extrapolation problems, we use the $^{56}$Ni mass and the ejecta mass derived from pseudo-bolometric light curves. Figure~\ref{fig:P16-19_pseudo_comp} shows the best-fit parameters for SN~2021bxu from the pseudo-bolometric light curve compared to a sample of SNe~IIb, Ib, Ic from \citet[][]{prentice16,prentice19}. In order to obtain the pseudo-bolometric light curve in the range $4000 - 10000$\,\AA, \citet[][]{prentice16,prentice19} make use of $BVRI$-bands and fit the \citetalias{arnett82} model to find the best-fit parameters, mainly $M_{\mathrm{Ni}}$ and $M_{\mathrm{ej}}$. SN~2021bxu shows the lowest value of the entire SE-SNe sample for $M_{\mathrm{Ni}}$ and $M_{\mathrm{ej}}$ as derived from the pseudo-bolometric light curves.

\begin{figure*}
	\includegraphics[width=\textwidth]{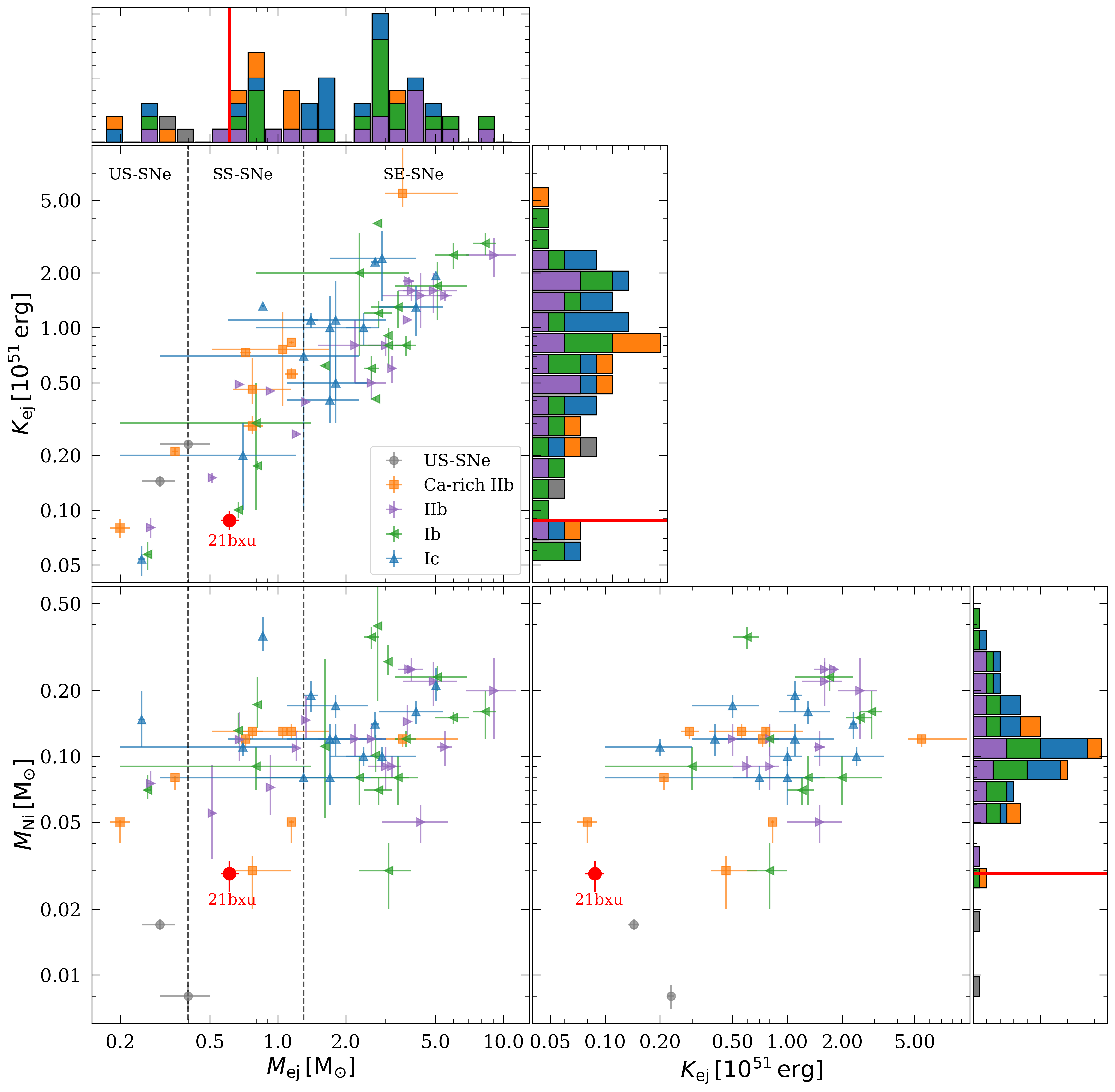}
    \caption{Comparison of SN~2021bxu with a sample of SNe~IIb, Ib, Ic from \citet[][]{taddia18} and \citet[][]{prentice16}, Ca-rich IIb SNe from \citet[][]{das22}, and US-SNe from \citet[][]{de18} and \citet[][]{yao20}. 
    The $^{56}$Ni mass ($M_{\mathrm{Ni}}$), ejecta mass ($M_{\mathrm{ej}}$), and kinetic energy ($K_{\mathrm{ej}}$) are derived using the full bolometric light curves for all SNe. The vertical dashed lines indicate the rough boundaries dividing ultra-stripped SNe (US-SNe), strongly-stripped SNe (SS-SNe), and stripped-envelope SNe \citep[SE-SNe;][]{das22}. The values for SN~2021bxu are shown in red in each panel.}
    \label{fig:MNi_Mej_Kej_D22}
\end{figure*}

Fully bolometric light curves are a better probe of the physical parameters than pseudo-bolometric light curves. Figure~\ref{fig:MNi_Mej_Kej_D22} shows the best-fit parameters from the bolometric light curve compared to a sample of SNe~IIb, Ib, Ic from \citet[][]{prentice16} and Ca-rich IIb SNe from \citet[][]{das22}. As seen in Figure~\ref{fig:MNi_Mej_Kej_D22}, $M_{\mathrm{Ni}}$ for SN~2021bxu is also the lowest in the bolometric sample of SE-SNe; the only lower values being that of the ultra-stripped SNe (US-SNe). $K_{\mathrm{ej}}$ for SN~2021bxu is also on the lower end of the comparison sample. $M_{\mathrm{ej}}$ for SN~2021bxu is closer to more common values and falls in the category of strongly-stripped SNe (SS-SNe). However, the best-fit parameters from a full bolometric light curve should be interpreted cautiously because the contribution to the total luminosity from the unobserved wavelengths is highly uncertain and treated differently by different studies. For example, \citet[][]{prentice16} assume a 10\% contribution from the unobserved wavelengths after considering $UBVRI$-bands and near-IR. In this study, we assume a blackbody SED for extrapolation to the unobserved wavelengths after direct integration in the observed bands.

Observed properties like low ejecta mass, low peak luminosity, faster-than-normal photometric and spectroscopic evolution are exhibited by the Ca-rich class of transients. They reach nebular phase quickly within as low as $\sim$1 month after explosion \citep[e.g.,][]{Shen19,nakaoka21,das22}. Core collapse Ca-rich transients are characterized by the presence of strong helium spectral features near peak and the appearance of calcium emission soon after. However, we do not find any strong calcium emission lines in the spectra, at least at the early epochs of $\sim$31 days after explosion. In SN~2021bxu, the presence of hydrogen along with neutron capture elements rules out most systems with WD progenitors.

The current popular theory suggests that Ca-rich SNe could result from low mass He stars ($M_{\mathrm{ZAMS}} \sim 8-12\,\mathrm{M_{\odot}}$) that are highly stripped down to less than $\sim 3\, \mathrm{M_{\odot}}$, suggesting a new class of SS-SNe \citep[e.g.,][]{das22,Ertini23}. SS-SNe form a transition class between SE-SNe and ultra-stripped SNe \citep[US-SNe;][]{Tauris15,das22}. As a result of their low initial masses, the resultant ejecta masses are also low (less than $\sim 1\, \mathrm{M_{\odot}}$) along with low $^{56}$Ni masses. \citet{Ertini23} showed that the progenitor scenario of the Ca-rich SN~2021gno can be modelled with a highly-stripped massive star with an ejecta mass of $0.8\,\mathrm{M_{\odot}}$ and a $^{56}$Ni mass of $0.024\,\mathrm{M_{\odot}}$, comparable to the values for SN~2021bxu.
 We see in Figure~\ref{fig:MNi_Mej_Kej_D22} that SN~2021bxu has $M_{\mathrm{ej}}$ consistent with the sample of Ca-rich IIb SNe and in the range demarcating SS-SNe, but $M_{\mathrm{Ni}}$ and $K_{\mathrm{ej}}$ are on the lowest end of the distribution.

Ca-rich IIb SNe are usually found in star-forming regions of galaxies and at smaller separations from the center of the host galaxy compared to Type I Ca-rich transients \citep[][]{das22}. As seen in Section~\ref{sec:host}, the host galaxy ESO~478-~G~006 is a star-forming galaxy with a star-formation rate of $33^{+16}_{-27}\, \mathrm{M_{\odot}\, yr^{-1}}$, typical of the hosts of Ca-rich IIb SNe \citep[][]{das22}. The projected physical offset of SN~2021bxu from its host is $9.2 \pm 0.6\,\mathrm{kpc}$, which is again typical for Ca-rich IIb SNe and SE-SNe \citep[][]{das22}. Despite its similarities with Ca-rich IIb SNe, the lack of nebular spectra of SN~2021bxu showing \CaII\,emission makes it difficult to conclusively judge whether SN~2021bxu is a Type IIb Ca-rich SN. Nevertheless, SN~2021bxu is an interesting object and shows that stars can explode with a small $M_{\mathrm{ej}}$ and $K_{\mathrm{ej}}$ with a large radius while making small amounts of $^{56}$Ni.

\section{Conclusion}
\label{sec:conclusion}
In this study, we present observations and analysis of a unique SE-SN, SN~2021bxu. As discussed in Section~\ref{sec:phot_anal}, we show that SN~2021bxu had a large initial decline in brightness followed by a short plateau not caused by H- or He-recombination, which is unusual for a Type IIb SN. It is on the faint end of the sample of SE-SNe from the literature, with the peak absolute magnitude of $M_r = -16.86 \pm 0.16 \, \mathrm{mag}$ and the average absolute magnitude during the plateau of $M_r \sim -15.9 \, \mathrm{mag}$. The pseudo-bolometric luminosity is also fainter compared to most other SE-SNe, with a peak of $\log(L_{\mathrm{pseudo}}/\mathrm{erg\,s^{-1}}) = 42.0$ and a distinct plateau phase at $\log(L_{\mathrm{pseudo}}/\mathrm{erg\,s^{-1}}) \sim 41.6$. The overall light curve shape in $gri$-bands matches most closely to that of Ca-rich IIb SNe, most of which show an initial decline and a second peak, and to that of SNe~IIb. The initial decline in the light curve of SN~2021bxu has a similar slope to the initial decline of SN~1993J but SN~1993J shows a distinct second peak from $^{56}$Ni at $\log(L_{\mathrm{pseudo}}/\mathrm{erg\,s^{-1}}) = 42.23$ and has a slower evolution at late-times indicating a larger ejecta mass.

With the presence of strong helium lines and weaker hydrogen lines, SN~2021bxu is a Type IIb SN (Section~\ref{sec:spec_anal}). We constrain the time of explosion using the \HeI\,$\lambda 5876$ line velocity and the blackbody radius evolution and find it to be $5.0 \pm 0.4\,\mathrm{days}$ before discovery. It evolves quickly to show absorption features from heavier metals like oxygen, calcium, silicon, iron, and neutron capture elements like barium and scandium, which get stronger over the 30-day spectral time-series. We note that SN~2021bxu shows spectral similarities to Type IIb SN~1993J as well as to Ca-rich IIb SNe during photometric phases with most of the same features observed. SN~1993J has velocities higher by a factor of $\sim$2 compared to SN~2021bxu. We also note similarities to SN~2021gno in terms of light curve evolution and modelling. From a photometric and spectroscopic analysis, we conclude that SN~2021bxu is a fast evolving SN, with a short distinct plateau phase not caused by H- or He-recombination and with some of the lowest observed line velocities compared to samples of Ca-rich IIb SNe and SE-SNe.

Following the modelling of the bolometric and pseudo-bolometric light curves in Section~\ref{sec:modelling}, we see that the light curves of SN~2021bxu can be well modelled by a composite model including interaction of the shock with an extended envelope of material surrounding the progenitor and the normal radioactive decay of $^{56}$Ni. We obtain the physical parameters for the explosion such as $M_{\mathrm{Ni}}$, $M_{\mathrm{ej}}$, $K_{\mathrm{ej}}$, and $T_0$, and the properties of the extended material $E_e$, $M_e$, and $R_e$ from bolometric and pseudo-bolometric light curves (see Table~\ref{tab:fit_params}). \citet{Ertini23} performed hydrodynamic modelling of SN~2021gno and similarly showed that the initial cooling phase can be explained by extended circumstellar material composed mainly of He, maybe with traces of H, and the second peak can be explained by the radioactive decay of $^{56}$Ni. We note that the $^{56}$Ni mass and kinetic energy for SN~2021bxu are on the lower end of the distribution for Ca-rich SNe and SE-SNe. The ejecta mass falls within the range of Ca-rich SNe and SS-SNe. 

Overall we determine that SN~2021bxu likely occurred from a lower-mass progenitor which had a large radius at the time of explosion and an extended envelope having experienced mass-loss potentially to a companion star, similar either to the SN~1993J scenario or to the strongly-stripped Ca-rich SNe such as SN~2021gno. However, to fully understand and characterize explosions similar to SN~2021bxu and to better constrain the parameters of the initial decline in an effort to constrain the immediate surroundings of the progenitor, further high precision multi-band observations of SNe in their infant stages are needed. POISE promises to deliver such a data set over the coming years. In addition to the early-time data, late-time data for such objects in the nebular phase would also prove beneficial in discerning if they are Ca-rich transients and allude to new classes such as SS-SNe.

\section*{Acknowledgements}
We thank the anonymous referee for their insightful and useful comments.
We thank Federica Chiti for helpful discussion. 

D.D.D. and B.J.S. acknowledge support from NSF grant AST-1908952.
C.A. and J.M.D. acknowledge support by NASA grant JWST-GO-02114.032-A and JWST-GO-02122.032-A. 
E.B. acknowledges support by NASA grant JWST-GO-02114.032-A.
L.G. acknowledges financial support from the Spanish Ministerio de Ciencia e Innovaci\'on (MCIN), the Agencia Estatal de Investigaci\'on (AEI) 10.13039/501100011033, and the European Social Fund (ESF) "Investing in your future" under the 2019 Ram\'on y Cajal program RYC2019-027683-I and the PID2020-115253GA-I00 HOSTFLOWS project, from Centro Superior de Investigaciones Cient\'ificas (CSIC) under the PIE project 20215AT016, and the program Unidad de Excelencia Mar\'ia de Maeztu CEX2020-001058-M.
M.D.S. and the Aarhus supernova group acknowledge support from the Independent Research Fund Denmark (IRFD, grant numbers 8021-00170B, 10.46540/2032-00022B) and the Villum Fonden (28021).
NUTS2 is supported in part by the Instrument Center for Danish Astrophysics (IDA).
J.P.A acknowledges funding from ANID, Millennium Science Initiative, ICN12\_009.
M.G. is supported by the EU Horizon 2020 research and innovation programme under grant agreement No 101004719.
T.E.M.B. acknowledges financial support from the 
Spanish Ministerio de Ciencia e Innovaci\'on (MCIN), the Agencia 
Estatal de Investigaci\'on (AEI) 10.13039/501100011033, and the 
European Union Next Generation EU/PRTR funds under the 2021 Juan de la 
Cierva program FJC2021-047124-I and the PID2020-115253GA-I00 HOSTFLOWS 
project, from Centro Superior de Investigaciones Cient\'ificas (CSIC) 
under the PIE project 20215AT016, and the program Unidad de Excelencia 
Mar\'ia de Maeztu CEX2020-001058-M.
M.N. is supported by the European Research Council (ERC) under the European Union’s Horizon 2020 research and innovation programme (grant agreement No.~948381) and by funding from the UK Space Agency.

Based on observations collected at the European Organisation for Astronomical Research in the Southern Hemisphere, Chile, as part of ePESSTO+ (the advanced Public ESO Spectroscopic Survey for Transient Objects Survey).
ePESSTO+ observations were obtained under ESO program IDs 106.216C and 108.220C (PI: Inserra).
Based on observations obtained at the international Gemini Observatory, a program of NSF’s NOIRLab, which is managed by the Association of Universities for Research in Astronomy (AURA) under a cooperative agreement with the National Science Foundation on behalf of the Gemini Observatory partnership: the National Science Foundation (United States), National Research Council (Canada), Agencia Nacional de Investigaci\'{o}n y Desarrollo (Chile), Ministerio de Ciencia, Tecnolog\'{i}a e Innovaci\'{o}n (Argentina), Minist\'{e}rio da Ci\^{e}ncia, Tecnologia, Inova\c{c}\~{o}es e Comunica\c{c}\~{o}es (Brazil), and Korea Astronomy and Space Science Institute (Republic of Korea).
This work was enabled by observations made from the Gemini North telescope, located within the Maunakea Science Reserve and adjacent to the summit of Maunakea. We are grateful for the privilege of observing the Universe from a place that is unique in both its astronomical quality and its cultural significance.


\section*{Data Availability}
The photometry presented in this paper is available in a machine-readable format from the online journal as supplementary material. A portion is shown in Tables~\ref{tab:bxu_phot}, \ref{tab:local_seq}, and \ref{tab:host_phot} for guidance regarding its form and content. The spectra presented in this paper are available via the WISeREP\footnote{\url{https://www.wiserep.org/}} archive \citep{Yaron12}.
 



\bibliographystyle{mnras}
\bibliography{sn2021bxu} 



\newpage
\appendix

\section{Photometry Data Tables} \label{app:phot_tables}
This section gives the photometry for SN~2021bxu in Table~\ref{tab:bxu_phot}, photometry of the local sequence stars used to calibrate the Swope light curves in Table~\ref{tab:local_seq}, and  photometry of the host galaxy ESO~478-~G~006 in Table~\ref{tab:host_phot}.

\begin{table*}
	\centering
    \caption{Photometry of SN~2021bxu}
	\label{tab:bxu_phot}
	\begin{tabular}{ccccc}
		\hline
		\hline
		MJD & Apparent & Absolute & Filter & Telescope \\
		(days) & Magnitude & Magnitude & & \\
		\hline
            59218.29	& $>22.29$	        	& $-$                   & $w$	& Pan-STARRS \\
            59245.12	& $>17.60$	        	& $-$                   & $g$	& ASAS-SN \\
            59245.26	& $>19.60$	        	& $-$                   & $o$	& ATLAS \\
            59251.06	& $17.17   \pm 0.09$	& $-17.16  \pm 0.19$	& $g$	& ASAS-SN \\
            59251.29	& $17.222 \pm 0.018$	& $-17.11  \pm 0.16$	& $c$	& ATLAS \\
            59252.03	& $17.462 \pm 0.007$	& $-16.86  \pm 0.16$	& $r$	& Swope \\
            59252.03	& $17.599 \pm 0.011$	& $-16.71  \pm 0.16$	& $i$	& Swope \\
            59252.04	& $17.315 \pm 0.011$	& $-17.03  \pm 0.16$	& $B$	& Swope \\
            59252.04	& $17.371 \pm 0.008$	& $-16.96  \pm 0.16$	& $g$	& Swope \\
            59252.04	& $17.372 \pm 0.010$	& $-16.95  \pm 0.16$	& $V$	& Swope \\
            59252.05	& $17.29   \pm 0.11$	& $-17.04  \pm 0.19$	& $g$	& ASAS-SN \\
            59253.03	& $17.689 \pm 0.015$	& $-16.63  \pm 0.16$	& $r$	& Swope \\
            59253.03	& $17.615 \pm 0.017$	& $-16.71  \pm 0.16$	& $V$	& Swope \\
            59253.03	& $17.672 \pm 0.012$	& $-16.66  \pm 0.16$	& $g$	& Swope \\
            59253.03	& $17.838 \pm 0.019$	& $-16.51  \pm 0.16$	& $u$	& Swope \\
            59253.04	& $17.657 \pm 0.019$	& $-16.68  \pm 0.16$	& $B$	& Swope \\
            59253.83	& $18.03   \pm 0.17$	& $-16.3   \pm 0.2$ 	& $g$	& ASAS-SN \\
		\hline
	\end{tabular}\\
	\begin{flushleft}
	\textbf{Note:} All magnitudes are given in the AB system. This table is available in its entirety in a machine-readable form in the online journal. A portion is shown here for guidance regarding its form and content.
	\end{flushleft}
\end{table*}

\begin{table*}
	\centering
    \caption{Photometry of the local sequence stars}
	\label{tab:local_seq}
        \resizebox{\textwidth}{!}{
	\begin{tabular}{cccccccc}
		\hline
		\hline
	      RA           & $\delta$      & $u$                  & $g$                   & $r$                  & $i$                   & $B$                  & $V$                  \\
            (J2000)      & (J2000)       & (mag)                & (mag)                 & (mag)                & (mag)                 & (mag)                & (mag)                \\
		\hline
            $32.386726$  & $-23.520622$  & $ 18.00 \pm  0.04$ & $ 15.592 \pm  0.008$  & $-                 $ &  $-                 $ & $ 16.035 \pm  0.010$ & $ 15.115 \pm  0.010$  \\
            $32.213863$  & $-23.394035$  & $ 16.20 \pm  0.02$ & $ 15.412 \pm  0.008$  & $ 15.157 \pm  0.012$ &  $-                 $ & $ 15.620 \pm  0.012$ & $ 15.267 \pm  0.010$  \\
            $32.385872$  & $-23.476055$  & $ 18.77 \pm  0.07$ & $ 16.238 \pm  0.006$  & $ 15.203 \pm  0.012$ &  $-                 $ & $ 16.697 \pm  0.012$ & $ 15.720 \pm  0.009$  \\
            $32.299397$  & $-23.491762$  & $ 16.92 \pm  0.02$ & $ 15.888 \pm  0.007$  & $ 15.498 \pm  0.008$ &  $ 15.365 \pm  0.005$ & $ 16.152 \pm  0.010$ & $ 15.659 \pm  0.009$  \\
            $32.245022$  & $-23.485291$  & $ 17.00 \pm  0.03$ & $ 15.969 \pm  0.007$  & $ 15.594 \pm  0.008$ &  $ 15.465 \pm  0.005$ & $ 16.223 \pm  0.010$ & $ 15.743 \pm  0.009$  \\
            $32.199989$  & $-23.418266$  & $ 17.87 \pm  0.04$ & $ 16.586 \pm  0.006$  & $ 16.047 \pm  0.008$ &  $ 15.853 \pm  0.005$ & $ 16.905 \pm  0.013$ & $ 16.294 \pm  0.010$  \\
            $32.345909$  & $-23.366911$  & $ 18.69 \pm  0.07$ & $ 17.580 \pm  0.007$  & $ 17.130 \pm  0.011$ &  $ 17.012 \pm  0.010$ & $ 17.87 \pm  0.02$ & $ 17.315 \pm  0.014$  \\
            $32.385616$  & $-23.374666$  & $-               $ & $ 18.673 \pm  0.017$  & $ 17.261 \pm  0.011$ &  $ 15.899 \pm  0.006$ & $ 19.17 \pm  0.05$ & $ 18.02 \pm  0.02$  \\
            $32.226913$  & $-23.380148$  & $-               $ & $ 18.568 \pm  0.015$  & $ 17.249 \pm  0.011$ &  $ 16.204 \pm  0.006$ & $ 19.23 \pm  0.06$ & $ 17.91 \pm  0.02$  \\
            $32.393101$  & $-23.397284$  & $ 18.62 \pm  0.06$ & $ 17.688 \pm  0.008$  & $ 17.266 \pm  0.011$ &  $ 17.132 \pm  0.011$ & $ 17.92 \pm  0.02$ & $ 17.442 \pm  0.015$  \\
            $32.283989$  & $-23.457907$  & $-               $ & $ 18.795 \pm  0.018$  & $ 17.449 \pm  0.012$ &  $ 16.290 \pm  0.006$ & $ 19.36 \pm  0.06$ & $ 18.10 \pm  0.02$  \\
            $32.393776$  & $-23.471512$  & $-               $ & $ 19.15 \pm  0.03$  & $ 17.874 \pm  0.016$ &  $ 17.387 \pm  0.013$ & $ 19.56 \pm  0.07$ & $ 18.55 \pm  0.04$  \\
            $32.342522$  & $-23.359810$  & $-               $ & $ 19.42 \pm  0.03$  & $ 18.034 \pm  0.018$ &  $ 16.771 \pm  0.008$ & $ 19.79 \pm  0.10$ & $ 18.71 \pm  0.04$  \\
            $32.308815$  & $-23.504641$  & $-               $ & $ 19.31 \pm  0.03$  & $ 18.14 \pm  0.02$ &  $ 17.499 \pm  0.015$ & $ 19.88 \pm  0.10$ & $ 18.65 \pm  0.04$  \\
            $32.287636$  & $-23.507380$  & $-               $ & $ 19.40 \pm  0.03$  & $ 18.46 \pm  0.03$ &  $ 18.01 \pm  0.02$ & $ 19.92 \pm  0.10$ & $ 18.87 \pm  0.05$  \\
            $32.300083$  & $-23.356009$  & $-               $ & $ 19.60 \pm  0.04$  & $ 18.39 \pm  0.02$ &  $ 17.487 \pm  0.015$ & $ 20.48 \pm  0.18$ & $ 19.07 \pm  0.06$  \\
            $32.279221$  & $-23.437853$  & $ 18.95 \pm  0.11$ & $ 18.96 \pm  0.02$  & $ 18.51 \pm  0.03$ &  $ 18.29 \pm  0.03$ & $ 19.15 \pm  0.05$ & $ 18.70 \pm  0.04$  \\
            $32.270084$  & $-23.510548$  & $ 19.74 \pm  0.16$ & $ 18.91 \pm  0.02$  & $ 18.66 \pm  0.03$ &  $ 18.57 \pm  0.04$ & $ 19.09 \pm  0.04$ & $ 18.76 \pm  0.04$  \\
            $32.232883$  & $-23.362158$  & $-               $ & $ 20.38 \pm  0.09$  & $ 19.30 \pm  0.06$ &  $ 18.52 \pm  0.04$ & $-                 $ & $ 19.87 \pm  0.13$  \\
            $32.285130$  & $-23.352280$  & $-               $ & $ 19.73 \pm  0.04$  & $ 19.46 \pm  0.07$ &  $ 19.12 \pm  0.08$ & $ 20.13 \pm  0.13$ & $ 19.61 \pm  0.11$  \\
            $32.226944$  & $-23.370340$  & $-               $ & $ 19.91 \pm  0.05$  & $ 19.42 \pm  0.06$ &  $ 19.17 \pm  0.08$ & $ 20.20 \pm  0.13$ & $ 19.71 \pm  0.12$  \\
            $32.355373$  & $-23.451790$  & $-               $ & $ 20.10 \pm  0.07$  & $ 18.78 \pm  0.03$ &  $ 18.24 \pm  0.03$ & $ 20.46 \pm  0.17$ & $ 19.13 \pm  0.07$  \\
            $32.256466$  & $-23.395746$  & $-               $ & $-               $  & $ 19.39 \pm  0.06$ &  $ 18.72 \pm  0.05$ & $-                 $ & $-                 $  \\
            $32.336956$  & $-23.473841$  & $-               $ & $ 20.22 \pm  0.07$  & $ 19.09 \pm  0.04$ &  $ 18.50 \pm  0.04$ & $-                 $ & $ 19.64 \pm  0.11$  \\
            $32.302109$  & $-23.528551$  & $-               $ & $ 20.86 \pm  0.15$  & $ 19.75 \pm  0.09$ &  $ 18.43 \pm  0.03$ & $-                 $ & $-                 $  \\
            $32.274239$  & $-23.415289$  & $-               $ & $ 20.78 \pm  0.14$  & $ 19.96 \pm  0.13$ &  $ 18.40 \pm  0.03$ & $-                 $ & $-                 $  \\
            $32.277119$  & $-23.527414$  & $-               $ & $-                 $  & $ 19.67 \pm  0.08$ &  $ 18.76 \pm  0.05$ & $-                 $ & $-                 $  \\
            $32.338875$  & $-23.392778$  & $ 19.78 \pm  0.17$ & $ 19.59 \pm  0.04$  & $ 18.98 \pm  0.04$ &  $ 18.65 \pm  0.04$ & $ 19.70 \pm  0.08$ & $ 19.25 \pm  0.07$  \\
            $32.276211$  & $-23.391331$  & $ 14.494 \pm  0.016$ & $-                 $  & $-                 $ &  $-                 $ & $-                 $ & $-                 $  \\
            $32.307014$  & $-23.431122$  & $ 16.41 \pm  0.02$ & $-                 $  & $-                 $ &  $-                 $ & $ 15.333 \pm  0.012$ & $-                 $  \\
            $32.352291$  & $-23.418312$  & $ 15.618 \pm  0.018$ & $-                 $  & $-                 $ &  $-                 $ & $ 14.742 \pm  0.012$ & $-                 $  \\
            $32.361607$  & $-23.421463$  & $ 15.777 \pm  0.018$ & $-                 $  & $-                 $ &  $-                 $ & $-                 $ & $-                 $  \\
		\hline
	\end{tabular}} \\
	\begin{flushleft}
	\textbf{Note:} All magnitudes are given in the natural Swope system. This table is available in its entirety in a machine-readable form in the online journal.
	\end{flushleft}
\end{table*}

\begin{table}
	\centering
    \caption{Photometry of the Host Galaxy ESO~478-~G~006}
	\label{tab:host_phot}
	\begin{tabular}{lcc}
		\hline
		\hline
		Filter & Magnitude & Mag. Error \\
		\hline
		Swift $UVW1$             & 14.96 & 0.02  \\
		Swift $U$                & 14.19 & 0.01  \\
		PS $g$                   & 13.05 & 0.01  \\ 
		PS $r$                   & 12.63 & 0.01  \\ 
		PS $i$                   & 12.34 & 0.01  \\ 
		PS $z$                   & 12.11 & 0.01  \\
		PS $Y$                   & 12.14 & 0.01  \\ 
		2MASS $J$                & 11.60 & 0.02  \\
		2MASS $H$                & 11.40 & 0.02  \\
		2MASS $K_{s}$ & 11.54 & 0.03  \\
		WISE $W1$                & 13.01 & 0.02  \\
		WISE $W2$                & 13.44 & 0.02  \\
		\hline
	\end{tabular}\\
	\begin{flushleft}
	\textbf{Note:} All magnitudes are given in the AB system. This table is available in its entirety in a machine-readable form in the online journal.
	\end{flushleft}
\end{table}


\label{lastpage}
\end{document}